\documentclass[12pt]{article}
\pdfoutput=1
\usepackage[a4paper, total={6in, 9in}]{geometry}
\usepackage{hyperref}

\usepackage{graphicx,subfig}
\usepackage{float}
\usepackage{verbatim}

\providecommand{\keywords}[1]
{
	\small	
	\textbf{\textit{Keywords---}} #1
}

\title{Triangular research and innovation collaborations in the European area}
\author{K. Angelou$^{1,2}$, M. Maragakis$^{1,2,3}$, K. Kosmidis$^{1,2}$, P. Argyrakis$^{1,2}$  \\
	\small $^{1}$Department of Physics, University of Thessaloniki - Thessaloniki, Greece \\
	\small $^{2}$Center of Complex Systems, University of Thessaloniki - Thessaloniki, Greece \\
	\small $^{3}$Department of Physics, International Hellenic University - Kavala, Greece
}
\date{} 

\begin{document}
\maketitle
\begin{abstract}
	In the current study, we examine the multiplex network of patents and European Framework Programmes (FPs) aiming to  uncover temporal variations in the formation patterns of triangles (a fully connected network between any three nodes). More specifically, the multiplex network consists of two layers whose nodes are the NUTS$2$ regions. On the first layer we depict the regions of the inventors that collaborated for the creation of a patent, and on the second those of the scientists in European Framework Programme (FP) funded projects. A link between two nodes exists when scientists or inventors from different regions collaborate. We split the network temporally into $28$ shorter sub-networks with a span of $6$ years each, and calculate the number of triangles formed at the end of the $6$-year period. Next, we shuffle the data creating again $28$ six-year randomized networks, in order to identify whether there is a hidden mechanism that favors a non-random behavior. Real and shuffled data are compared using a z-score, a measure of the differences of standard deviations between them. In addition, we repeat the same analysis using the clustering coefficient, which is the number of triangles over the number of triples (possible triangles). The results show that triangular FP collaborations tend to be favored over random ones, while in patents the case is strongly the opposite. Furthermore, results using triangles tend to be more comprehensive as opposed to those of the clustering coefficient. Finally, we identify which NUTS$2$ regions frequently exhibit a high clustering coefficient in either of the layers, and we present a map with these values for all regions. The results of this research can help policy making organizations understand the spatial dimension of subsidized research and patented innovation collaboration networks.
\end{abstract} \hspace{10pt}


\section{Introduction}\label{intro}
Collaboration networks have been studied for many years, and as their name implies, they deal with the interactions between different types of actors in socially based systems \cite{Wasserman1994a, Scott1991, Borgatti2009}. The analysis of such networks is important, since it helps, for example, the study of relationships between individuals, extract patterns in the network's formation, or identify the most influential nodes. Newman studied the structure \cite{Newman2001} and characteristics of scientific collaboration networks \cite{Newman2001a}.
In research, collaborations have been studied in the past by sector \cite{Broekel2015}, by actor-based differences \cite{Broekel2012, Tomasello2014}, or by geographic origin \cite{Scherngell2009}. Such studies focus mainly on the structure \cite{Abbasi2012} and evolution \cite{Ding2011} of the networks.\\
In innovation, efforts to study the structure of the patent network have attracted much interest \cite{Fleming2007,Choe2017}, with the spatial aspect gathering enough focus \cite{Maggioni2011, Sun2016}, while several studies try to predict emerging technologies \cite{Erdi2013,You2017}.\\
Apart from studying single layer networks, it is also vital to study multilayer networks \cite{Kivela2013,Aleta2019} in order to find out how the evolution of a specific network affects the other. The last decade there has been renewed interest in multiplex social networks \cite{Torenvlied2018}, although sociologists had introduced the multilayer perspective in social relationships during the late '$30$s  \cite{RoethlisbergerF.J.;DicksonW.J.;WrightHaroldA.;Pforzheimer1939,Verbrugge1979}. More recently, Mittal et al \cite{Mittal2018} proposed a new way of calculating closeness centrality  for multiplex social networks, and Hosni et al \cite{Hosni2020} examined rumor propagation while trying to reduce its influence. Ansari et al \cite{Ansari2011} tried to model the various types of relationships between actors and Jalili et al \cite{Jalili2017} wanted to predict future links in the multiplex network of Twitter and Foursquare. Ramezanian et al \cite{Ramezanian2015} addressed the diffusion problem on multiplex social networks by extending a game-theoretic diffusion model, while Nguyen et al \cite{Nguyen2015} proposed two ways of applying community detection in multiplex social networks. In research and innovation, research is mostly limited in paper-patent studies \cite{Li2015}, and does not generally use the notion of multiplex networks, but rather compares the separate layers characteristics \cite{Magerman2015, Wanzenbock2014,VanderPol2018,Landini2015}. To the best of our knowledge similar work on multiplex networks of knowledge and innovation has only been done in \cite{DeStefano2013} where this approach was used to analyze the interactions among authors and inventors in the region of Trieste, Italy.\\
Triangles, essentially a trio of nodes that are all connected to each other, is a notion that dates back to $1908$ \cite{Simmel1908,Kuper1951}, and is used to study the structure of a multiplex collaboration network. Triangles have been used in various network studies and their most notable application is on structural analysis. Lambiotte et al. \cite{Lambiotte2008} use triangles in a mobile phone communication network in order to perform geographic analysis and identify inherent communities, as well as study the network's cohesion. Zhang et al \cite{Zhang2020} study the structure of landscape patterns in a coal mine area and make use of a triangle-based metric in order to examine the network stability. Antal et al \cite{Antal2006} study the dynamics of a social network by examining its balance based on triads and Dimitrova et al \cite{Dimitrova2020} use triangles as part of their multiplex networks structure analysis.\\
The main aim of the current study is to uncover temporal variations in the formation patterns of triangles in the networks of patents and European Framework Programmes (FPs) separately, as well as their common ones (triangles present in both layers). We focus on them because they are an indicator of whether there is a preference in a network to form three-way collaborations, which simple link analysis cannot unveil. Our results can prove valuable for policy-making organizations, ministries or other funding authorities, as they provide hints in the existing structure of the subsidized research and patented innovation landscape of the entire European area. We compare our results to the networks' local clustering coefficient \cite{Watts1998,Holland1971}, a well known metric used for studying the structure of social-collaborative networks.\\
The paper is structured as follows. Section \ref{data} presents the data used for our study, section \ref{methodology} describes the steps followed for the analysis, section \ref{results} contains the results obtained, and finally, section \ref{conclusions} draws the conclusions of this work.\\

\section{Data}\label{data}
The multiplex network that we study consists of two layers, patent and scientific collaborations. Our aim is to identify the evolution of collaborations in research and innovation between different regions. However, the data at the individual scientist level are quite sparse and, instead, we use their "Nomenclature of Territorial Units for Statistics" (NUTS) region codes. As a result, the nodes of our networks are the NUTS$2$ regions of the scientists/patent creators origin. The NUTS$2$ methodology forces the division of a country into smaller regions of typically about $800,000$ to $3,000,000$ population. In our network approach, links between two individual NUTS$2$ codes are introduced when scientists that originate from two different regions collaborate for the creation of a patent or in a European Framework Programme (FP$5$-$7$, and Horizon$2020$) project.\\
The first layer of the multiplex network is constructed using data from subsidized research, namely from the FP$5$-$7$ and Horizon$2020$ Framework Programme projects. Although FP$5$ started in $1998$, we begin our study from the year $2000$ because data prior to that year are quite sparse, and could prove misleading. This also defines the maximally available time period for the study of the multiplex network. FP data have been extracted by the Community Research and Development Information Service (CORDIS) and the aggregate network contains in total $34,061$ projects, $330$ different regions and $40,032$ unique links.\\
The contents of the patent layer have been derived by the European Patent Office (EPO) and contain patents for the years $1978$ to $2020$. However, and in order to be in compliance with the FP database, we utilize only patents registered since $2000$ and onwards. The total number of patents that have been registered in any NUTS$2$ region by at least two inventors since then is $48,466$. The maximum number of regions participating in the aggregate network is equal to $323$, while the unique links are $5,692$ in total.\\
It should be noted here that while the number of patents is larger than the number of projects, the number of unique links in the patent layer is less than the number of unique links in the projects layer. This is an indication that the two layers have differences in their connectivity. Indeed, given that the number of unique links is much smaller in the patents than in the FPs, one expects that this layer is less dense that the FP one.\\
As analyzed in the methodology the multiplex network is studied in smaller time periods, and obviously, for each time period the number of nodes and links is smaller than the total one. On average the number of nodes is $291$ for patents and $324$ for FP, while the average number of links is $2,669$ and $27,948$, respectively.\\
Both databases contain geographic information about the scientists/inventors location. However, at some occasions the NUTS$2$ codes had to be extracted manually and were not listed as a separate field. The databases also contain a field that lists the duration of the patents and the FP projects. It should be noted that about $20\%$ of the links' duration data are missing here for the patent layer (although such data in some cases may be non-existent because the patent protection still exists), while for FP projects practically almost all collaborations have a "death" (link removal) time listed. This information is used for the temporal analysis in order to include the death of a link, as a more realistic approach.\\

\section{Methodology}\label{methodology}

\begin{figure}[!ht]
	\begin{center}
		\subfloat[\label{fig:fig1a}]{{\includegraphics[width=0.33\textwidth]{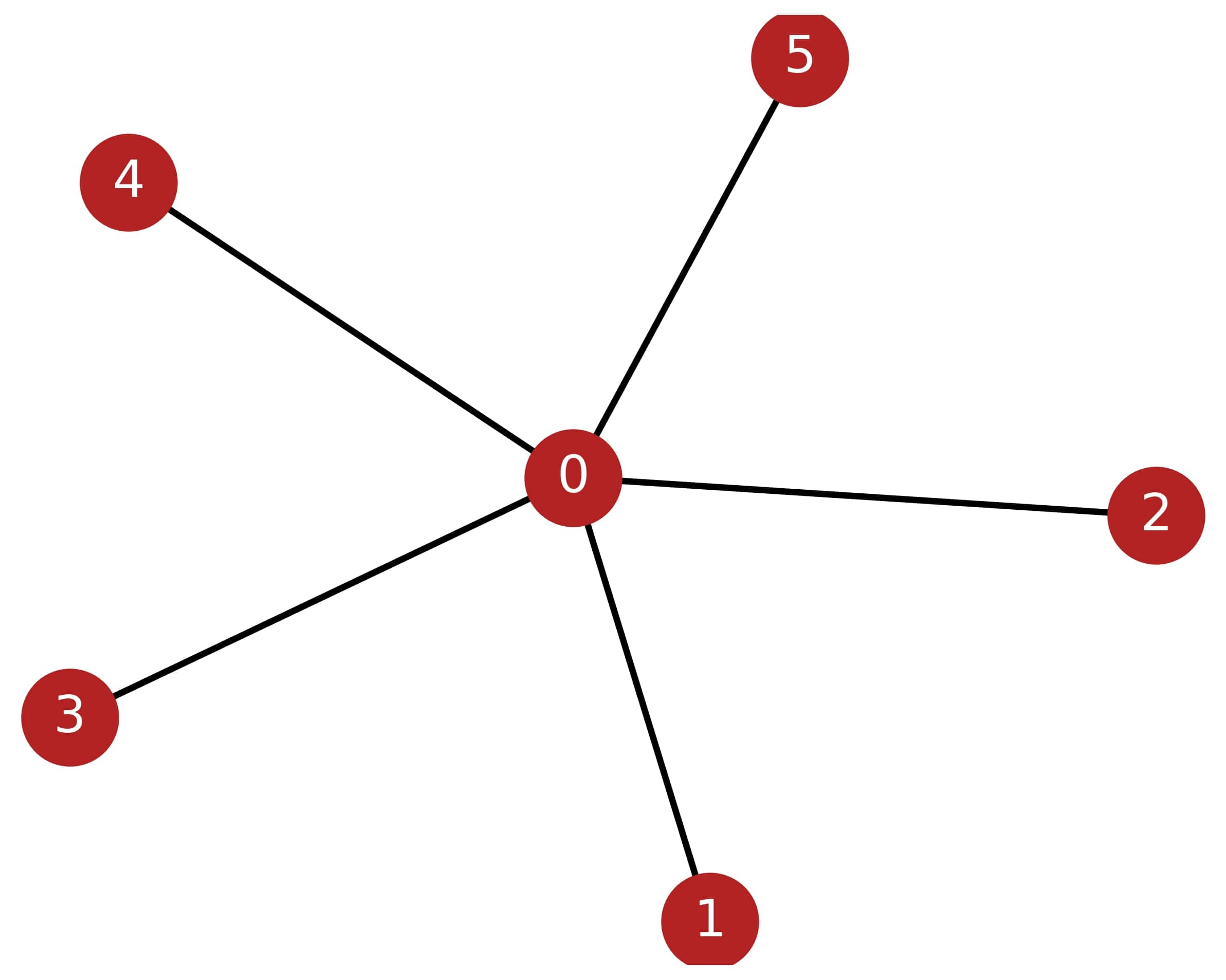} }}
		\subfloat[\label{fig:fig1b}]{{\includegraphics[width=0.33\textwidth]{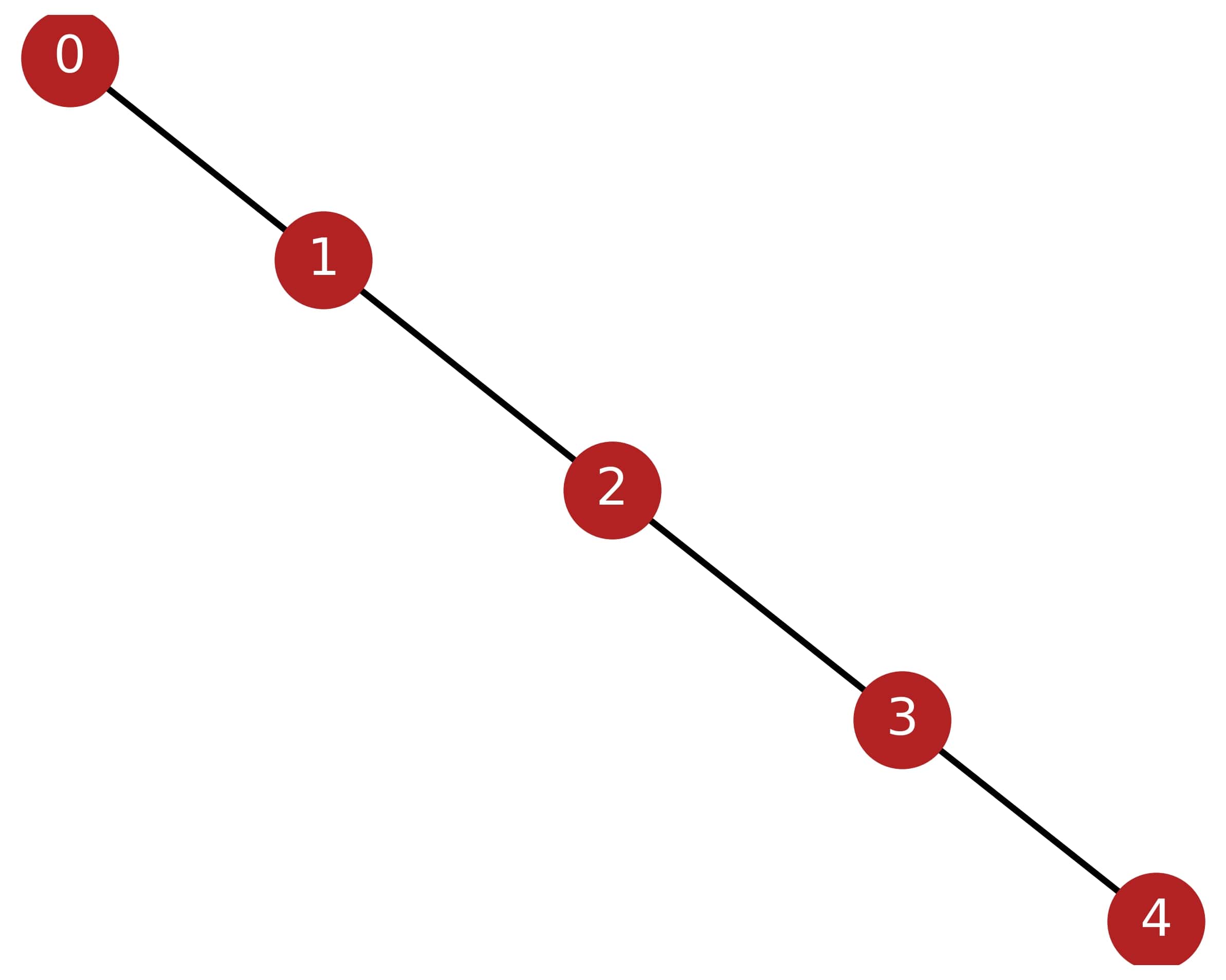} }} 
		\subfloat[\label{fig:fig1c}]{{\includegraphics[width=0.33\textwidth]{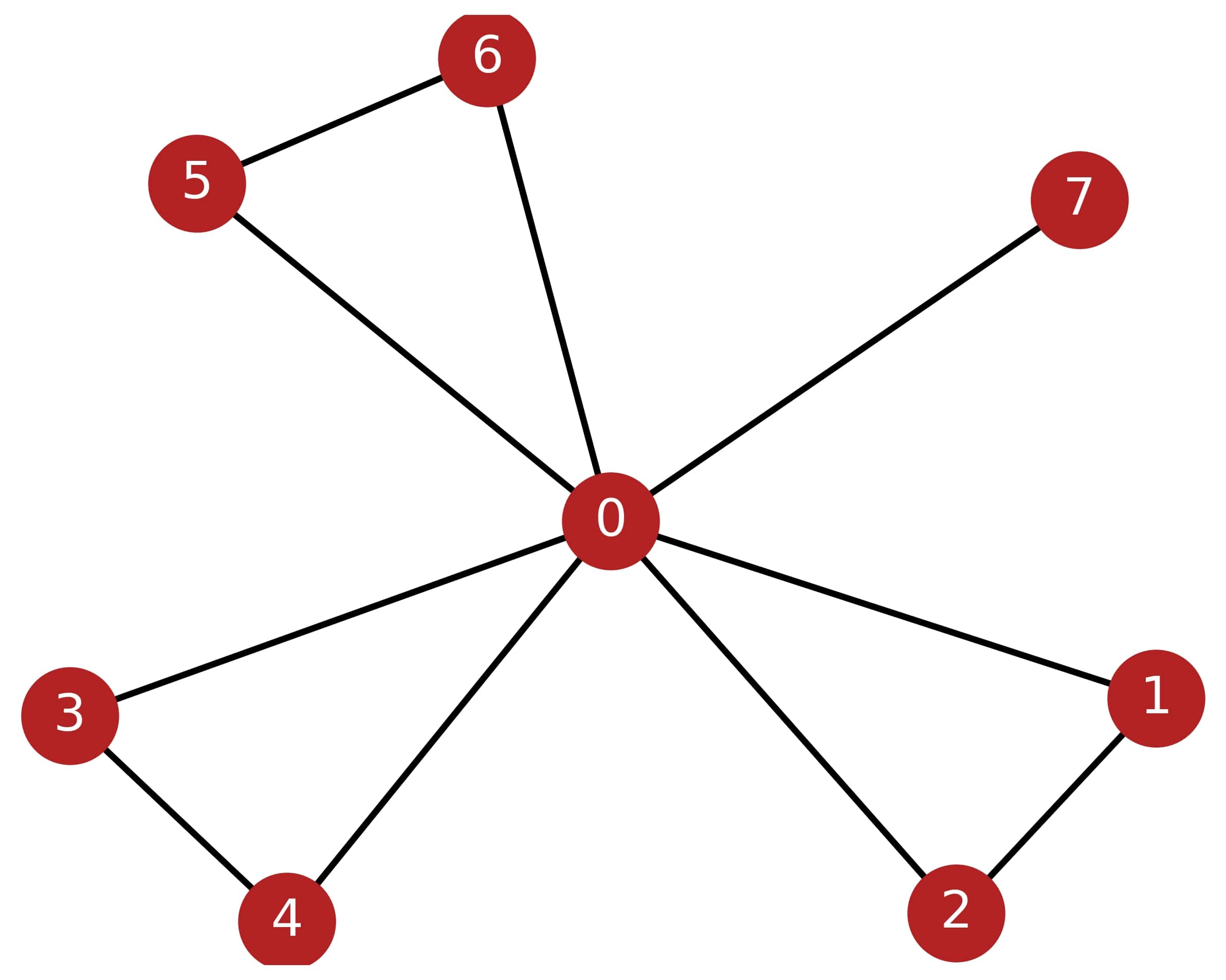} }}
		\caption{Schematic of 3 networks with various types of collaboration between the nodes. 
			a) Star network, where node $0$ may be connected to many other nodes, however, these are all isolated from each other and, thus, its local clustering coefficient is $0$ (same for nodes $1-5$). 
			b) Linear network, where all the nodes have local clustering coefficient $0$. 
			c) Triangles are formed in this type of network, so all nodes have a local clustering coefficient greater than $0$, apart for node $7$ which does not participate in any triangle. More specifically, nodes $1-6$ clustering coefficient is $1$, and node's $0$ is $0.14$.}
		\label{fig:fig1}
	\end{center}
\end{figure}

As mentioned in section \ref{intro}, triangles have been used in various studies for the structural analysis of networks. Our focus is on identifying collaboration triangles that exist in the multiplex network over long periods of time and, thus, regions that may have persistent collaboration patterns. We also seek to find those time points where structural changes lead to the significant increase, or decrease of such forms of inter-regional collaborations.\\
Figure \ref{fig:fig1} depicts some characteristic network cases that correspond to different collaboration patterns. Fig. \ref{fig:fig1a} shows a star-like network where the central node represents a strongly influential region with which all other regions prefer to connect to. Such a structure does not contain any triangles, the peripheral regions are isolated. Indeed, according to the definition of the local clustering coefficient \cite{Watts1998} of node $i$:

$$C_i = \frac{\lambda_G(v)}{\tau_G(v)}$$

where $\lambda_G(v)$ is the number of triangles, one of which is $v$, and $\tau_G(v)$ is the number of triples ($2$ links and $3$ nodes), where node $v$ is incident to both edges, the resulting value for the central node is $0$ and, all other nodes also exhibit a $C_i$ value of $0$. The same applies to a linear network, fig. \ref{fig:fig1b}, where no nodes participate in any triangle and $C_i$ is again $0$ for all nodes. In contrast to these very specific types of collaboration, there can be a star-like network, whose nodes are not isolated but are connected to each other, fig. \ref{fig:fig1c}. In this type of structures the nodes are more efficiently connected. This is shown by their local clustering coefficient, which is $1$ for all nodes, except for $0$ ($C_0=0.14$) and $7$ ($C_7=0$). An important limiting case is that of a fully connected network where all nodes connect to all other nodes (not shown schematically). In this case the local clustering coefficients would be $C_i=1$ for all $i$.
In our study, we want to quantitatively find out which type of collaboration and how often a triangle is met in a multiplex research and innovation collaboration network, as opposed to the case of a similar randomized multiplex network. This helps us to classify qualitatively the network topology. Such an approach has also been used in the study of research and innovation networks in \cite{Tahmooresnejad2018}.\\

\begin{figure}[!ht]
	\begin{center}
		\subfloat[\label{fig:fig2a}]{{\includegraphics[width=0.35\textwidth]{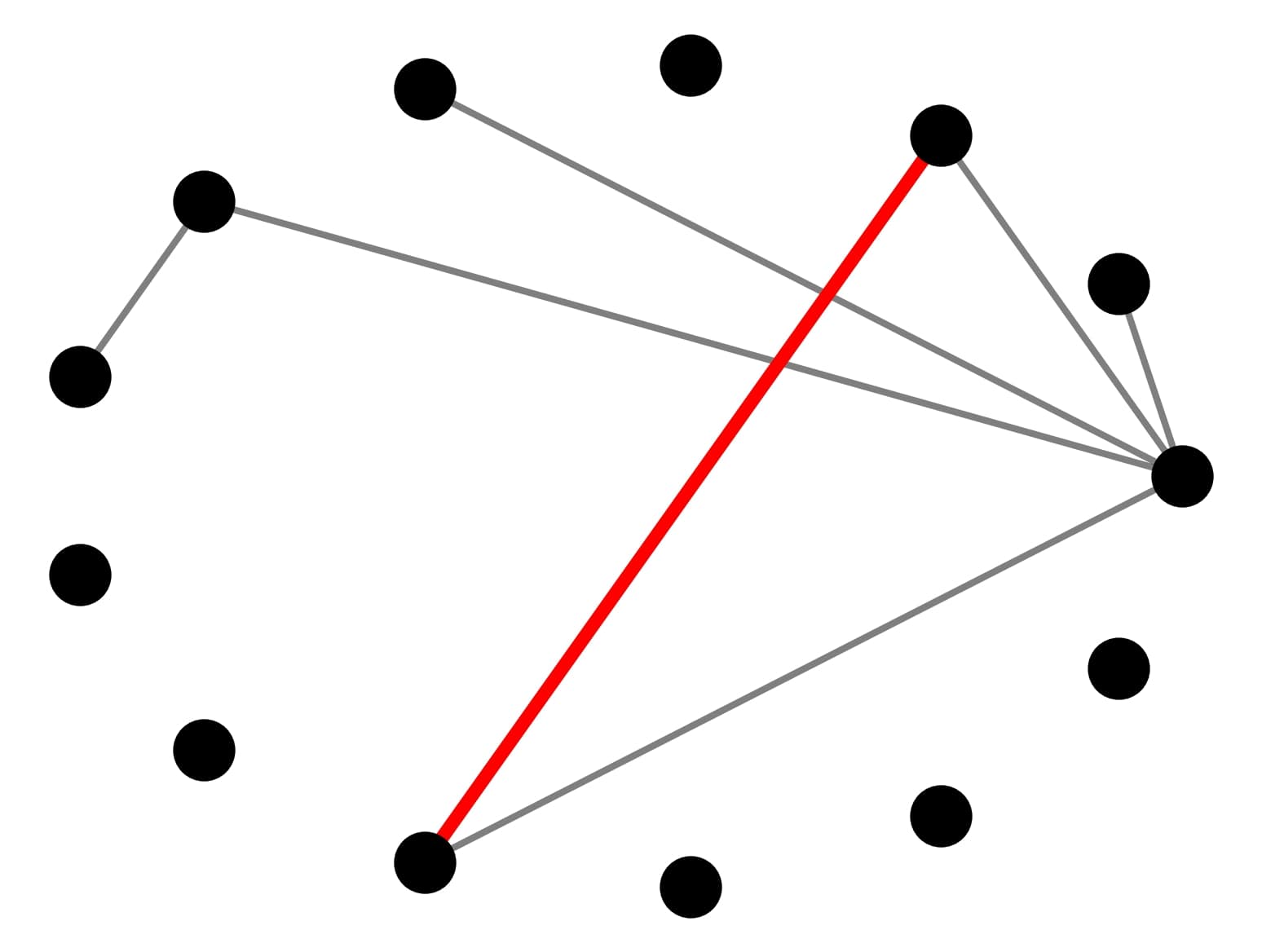} }}
		\subfloat[\label{fig:fig2b}]{{\includegraphics[width=0.35\textwidth]{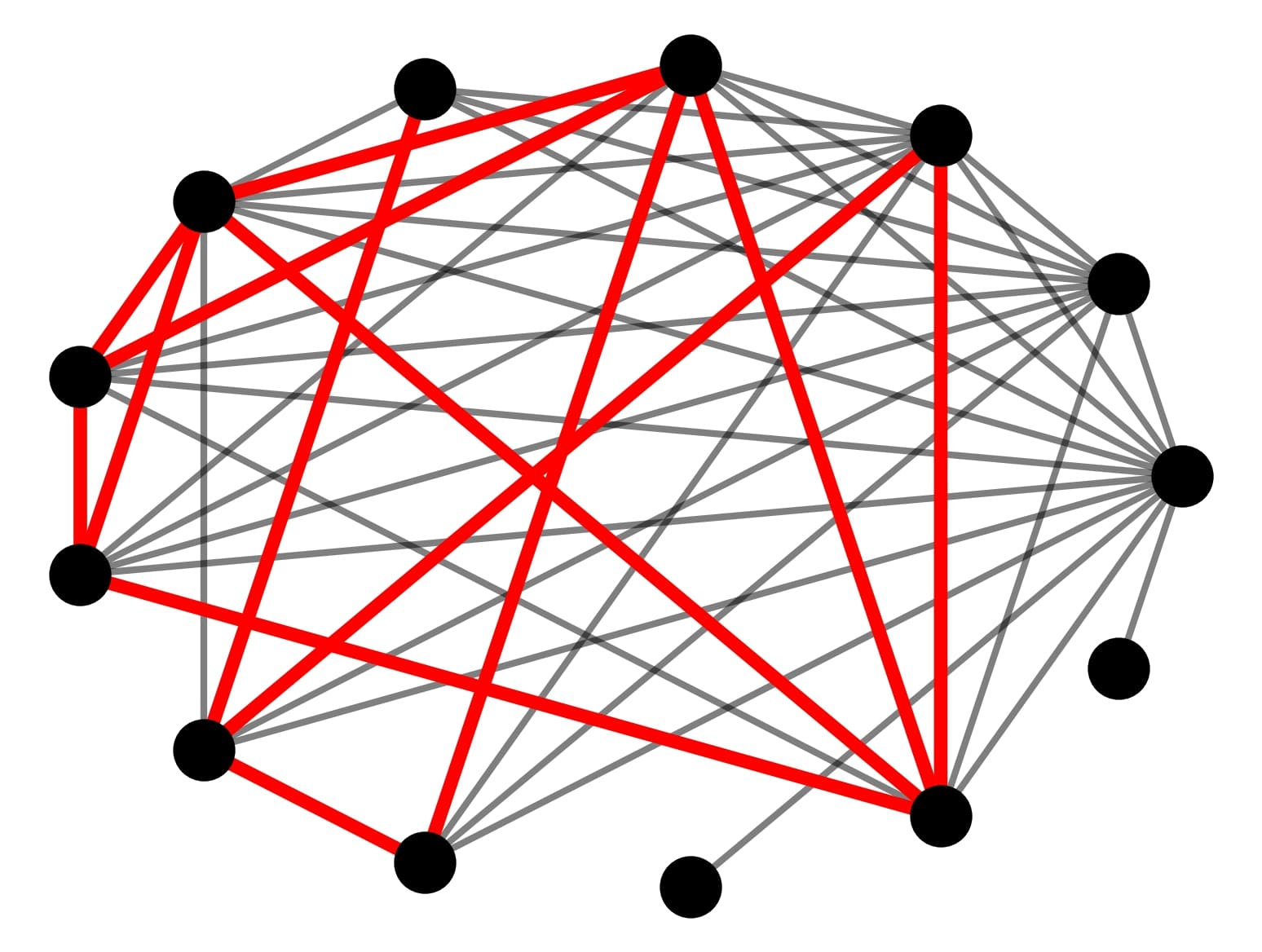} }} \\
		\subfloat[\label{fig:fig2c}]{{\includegraphics[width=0.35\textwidth]{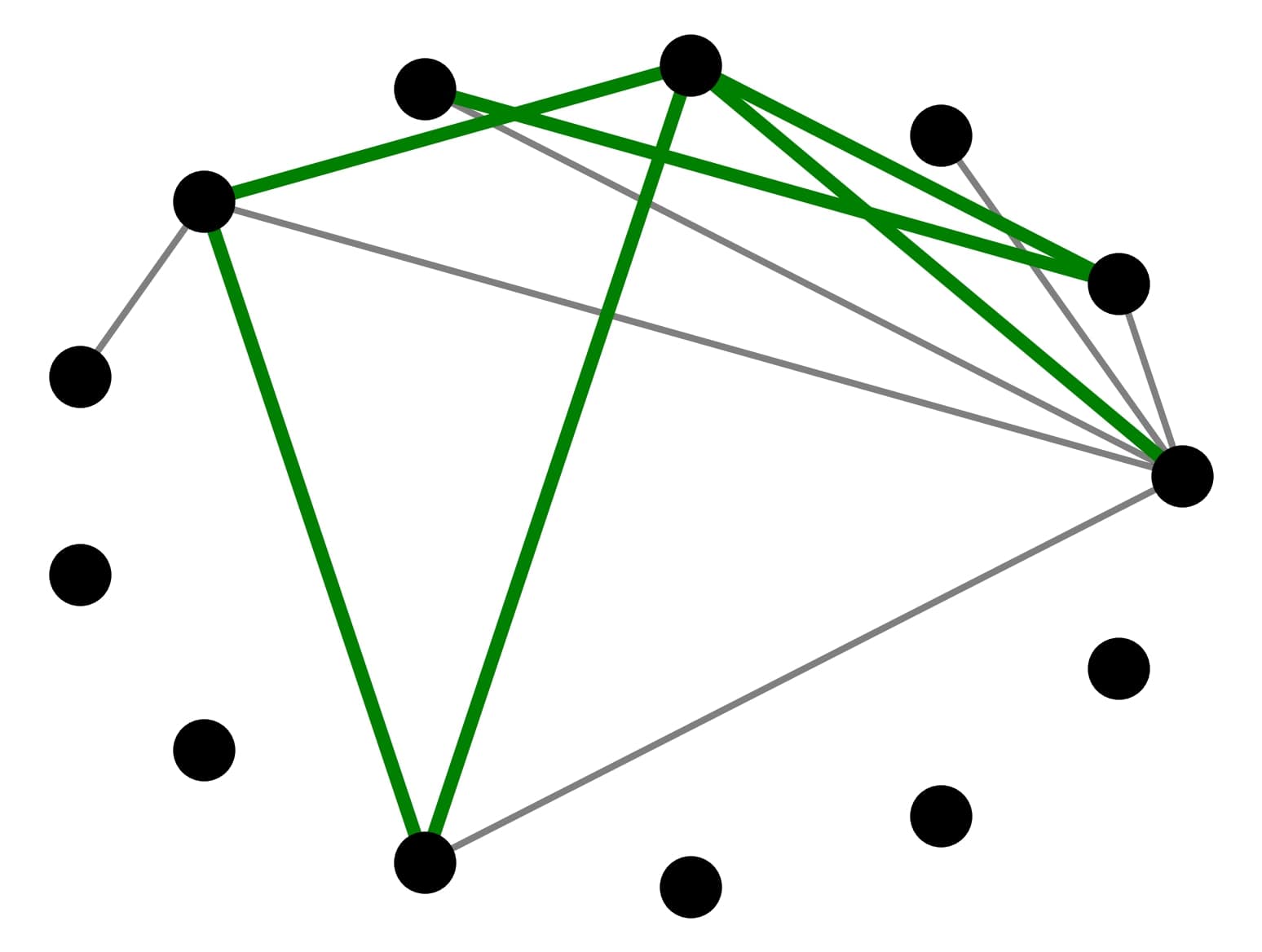} }}
		\subfloat[\label{fig:fig2d}]{{\includegraphics[width=0.35\textwidth]{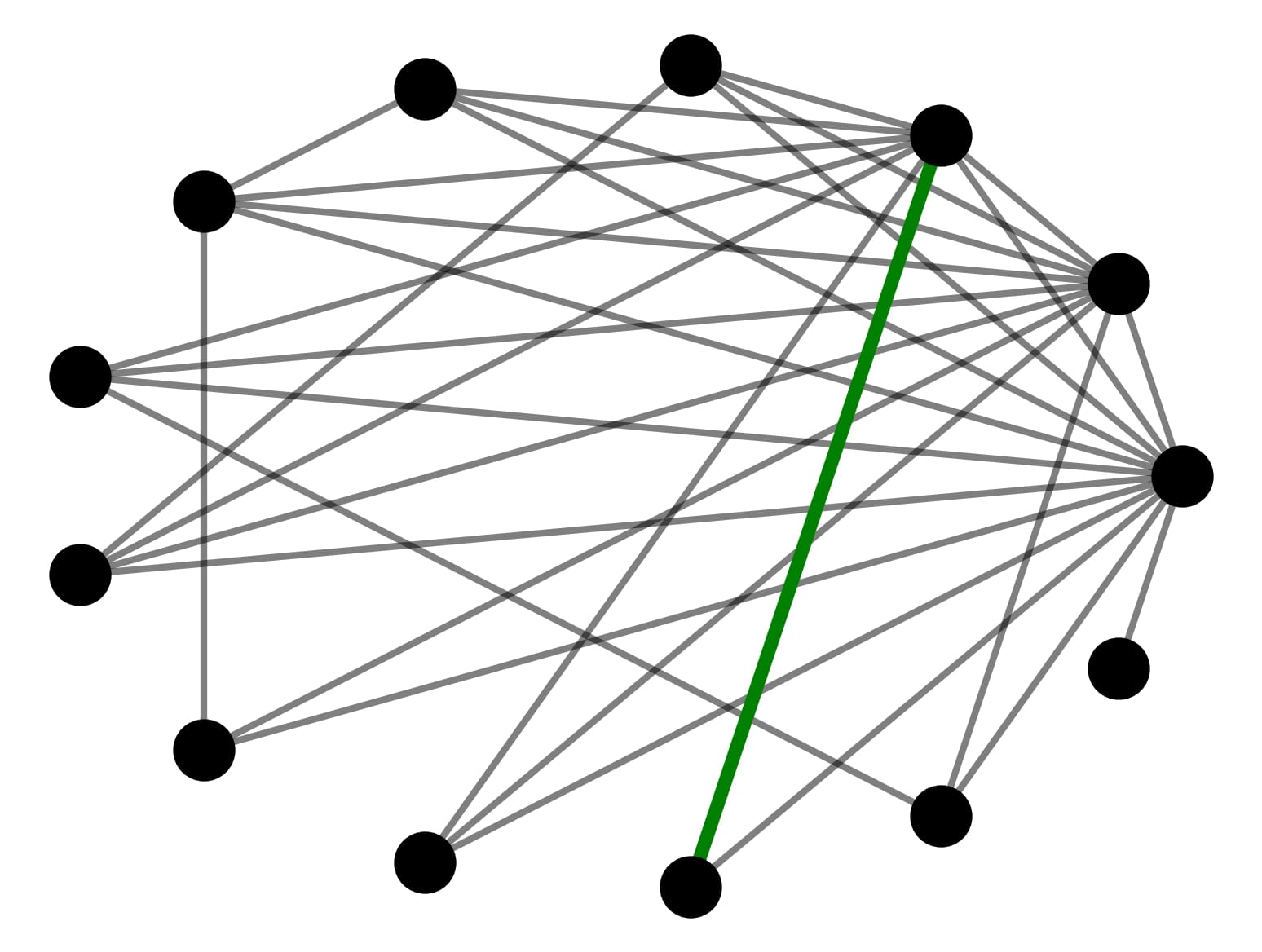} }}
		\caption{
			Schematic of the evolution of a multiplex network with temporal data and links that "die" at some point. Figures a) and b) are the two layers of the multiplex network that have grown until a specific date, and figures c) and d) are the same layers at a later point of time. Red lines indicate the links that are about to be removed, while green lines are the new links that have been inserted into the network.
		}
		\label{fig:fig2}
	\end{center}
\end{figure}

We use the sliding windows method \cite{Datar2002,Angelou2020,Angelou2020a} that allows for the division of an evolving network into smaller networks. To be more specific, each sub-network initiates at the beginning of every year (January $1$st) or in the middle of it (July $1$st) for all years since $2000$, and up to $2014$. We allow for a growth period of exactly $6$ years for the sub-networks to reach a relative plateau in their evolution. At each date that patents or FP projects are registered, links are inserted into the multiplex networks, or removed if a collaboration has reached the "death" date. Figure \ref{fig:fig2} shows an illustration of the patent and the FP layer, at two different points of time, that links have been removed or inserted. At the end of each $6$-year period the number of existing triangles is calculated. In addition, we calculate the local clustering coefficient $C_i$ of each node, and subsequently the average value of the local clustering coefficient, $\overline{C_i}$, for all nodes, $i$. We , thus, make a qualitative comparison between the results of the triangles approach and the clustering coefficient one.\\
We repeat the entire analysis for randomly shuffled networks of both layers. This is done in order to find out if there is an underlying mechanism responsible for any observations, or whether such results can occur by chance. More specifically, we shuffle the links while the degree distribution remains the same in both layers and reconstruct the multiplex network. We take care that the number of projects/patents per sliding window remains the same. However, we shuffle the dates that projects and patents are inserted into the network so as to ensure greater randomization, even during the networks' evolution.\\
In order to compare the results between the real and the shuffled data we use the standard score or z-score \cite{Spiegel2017}, given by:
$$ z = \frac{x-\overline{x}}{s}$$
where $x$ is the real value, $\overline{x}$ is the average value of the shuffled data, and $s$ their standard deviation. What this metric does is to calculate how many standard deviations the real data differ from the mean shuffled data. Randomly occurring networks with no preferential attachment in the creation process would have produced a z-score value around $0$. The comparison will take place both in triangles and the local clustering coefficient analysis.\\

\section{Results and discussion}\label{results}

As mentioned in section \ref{methodology}, the sliding window method creates $28$ multiplex sub-networks, i.e. $28$ patent sub-networks and $28$ FP sub-networks. For each set of sub-networks we identify and calculate the patent triangles and the FP triangles. 
We then remove from each layer their common links, those existing between the same two nodes in both layers, and place them into a new network, the common multiplex one.

\begin{figure}[!ht]
	\begin{center}
		\subfloat[\label{fig:fig3a}]{{\includegraphics[width=0.43\textwidth]{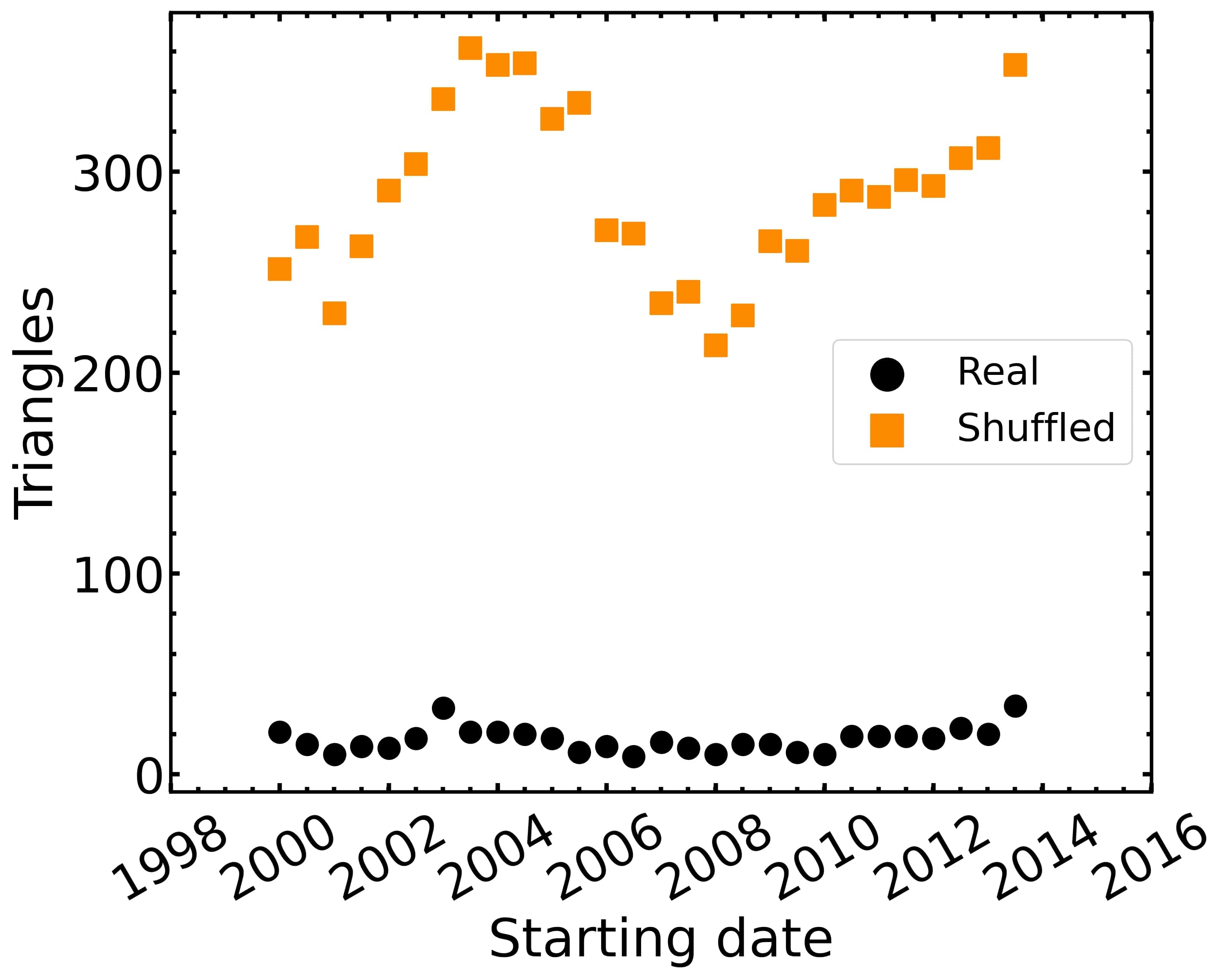} }}
		\subfloat[\label{fig:fig3b}]{{\includegraphics[width=0.43\textwidth]{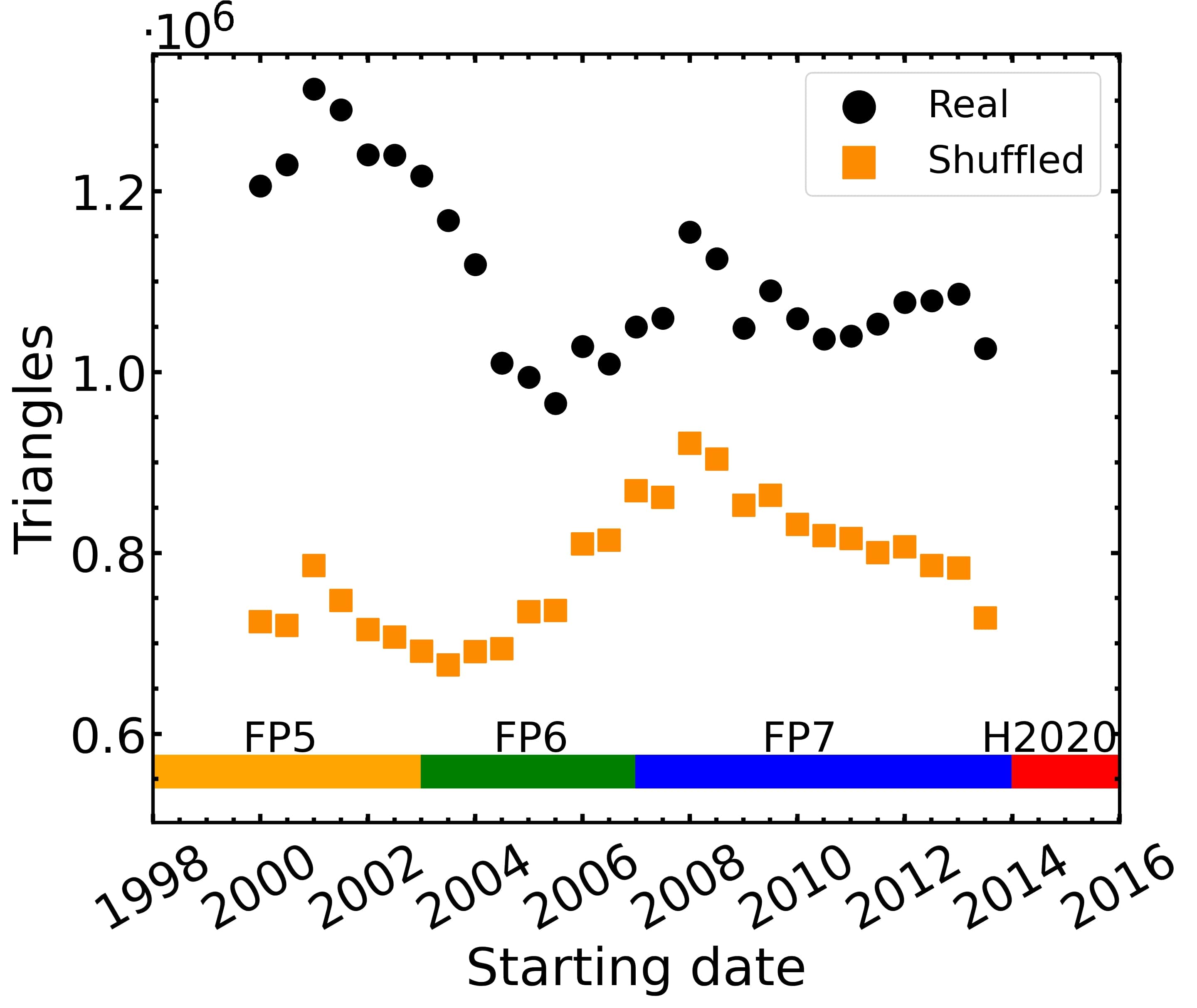} }} \\
		\subfloat[\label{fig:fig3c}]{{\includegraphics[width=0.43\textwidth]{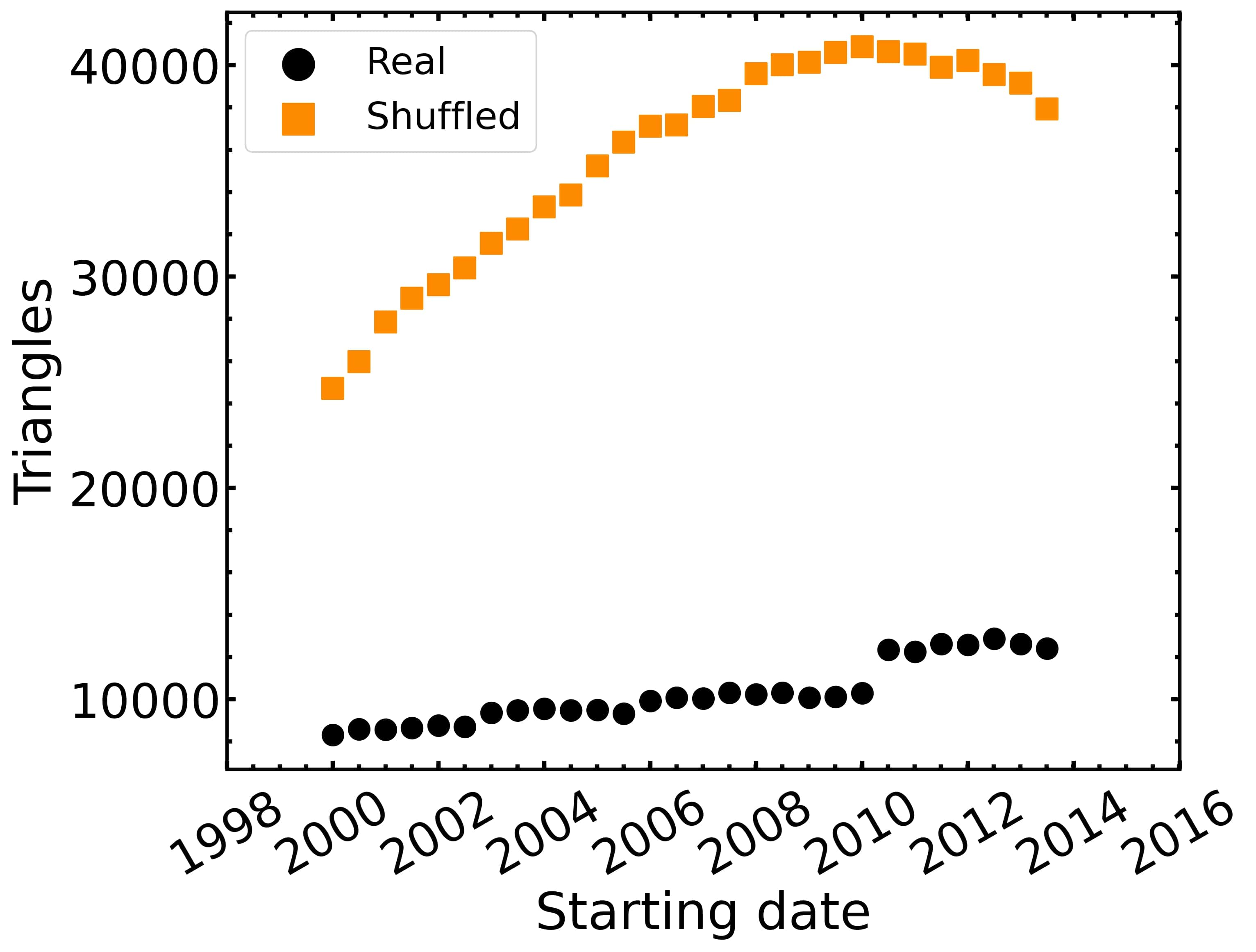} }}
		\subfloat[\label{fig:fig3d}]{{\includegraphics[width=0.40\textwidth]{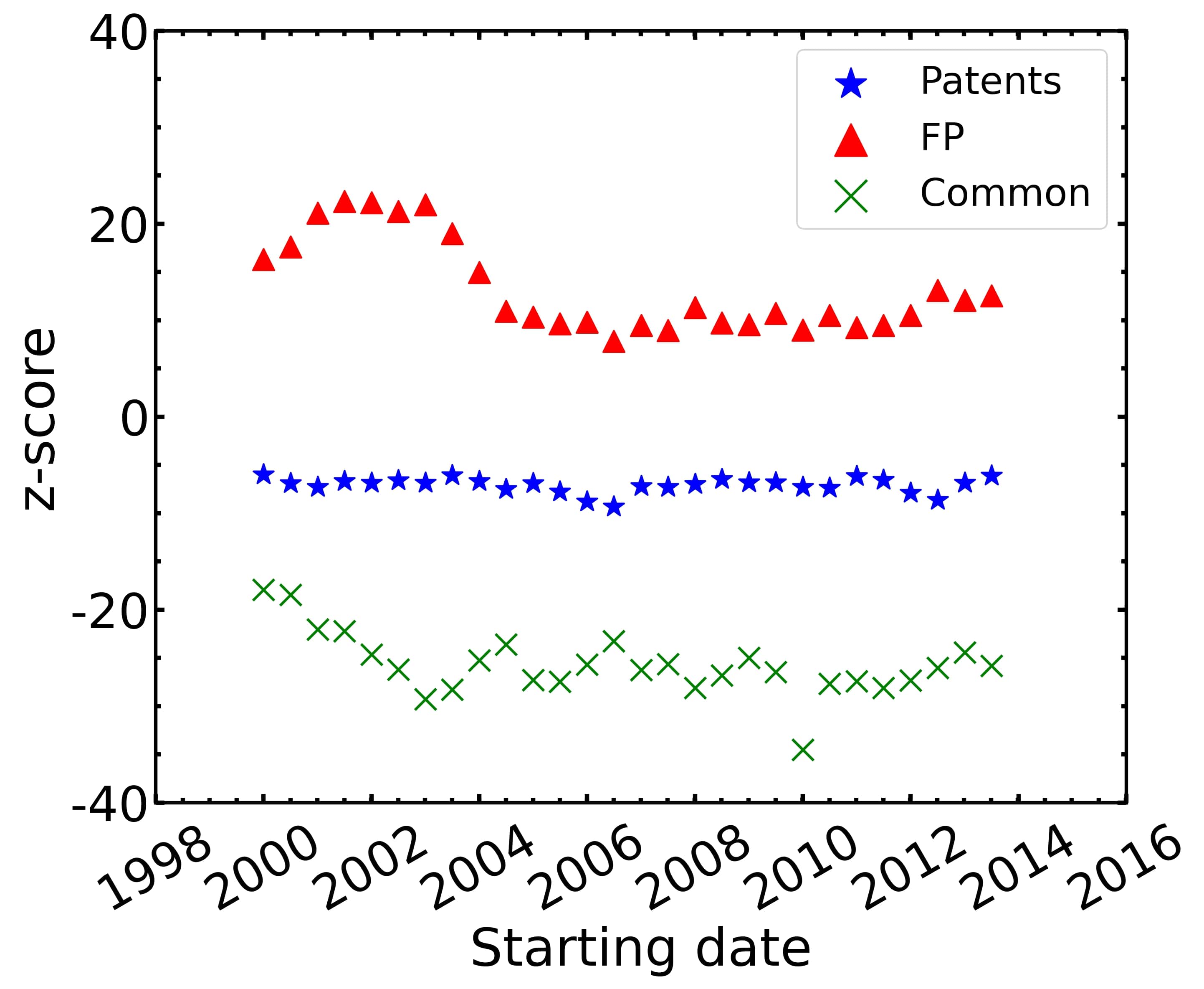} }}
		\caption{(Color online) The number of triangles at the end of the $28$ windows, versus their starting dates for a) patents, b) FP, and c) their common network. Black dots represent the real data, and orange squares represent the average of $50$ shuffled networks. The rectangular colored areas (colors are random) at the bottom of figure b show the FPs duration of each. d) z-score of the number of triangles for the patent (stars), FP (triangles) and Common (cross) networks versus time.
		}
		\label{fig:fig3}
	\end{center}
\end{figure}

We follow the same procedure for the $50$ shuffled networks that we have created for each of the $28$ real sub-networks. The results show that the real world patent layer, fig. \ref{fig:fig3a}, and the common network, fig. \ref{fig:fig3c}, are much more likely to form more isolated collaborations (simple links) when compared to their shuffled versions. In fact, shuffled data tend to form many more triangles (up to about $10$ times more in patents and $4$ in the common network) for the entire time period studied. 
On the other hand, real FP data tend to form more triangles than their corresponding shuffled data, fig. \ref{fig:fig3b}. This effect is possibly due to the specifics of EU funding rules, which promote in most funding calls a collaboration of three or more partners. Thus, triangular collaborations are favored and successful ones are most often those with even more than $3$ partners. The z-score values of all three cases prove that these results are not random as they range far from $3$ typical standard deviations. In fact the patents and the common network, fig. \ref{fig:fig3d}, show negative z-scores and the FP positive ones, agreeing with the observations of the previous plots. This behavior is, therefore, non random and hides an inherent preference in the network creation process. 

It is worth mentioning that the FP layer results show a much higher z-score up until $2004$, than they do until the end of our datasets. This may be related to changes in the mechanism of the FP layer creation process, namely the end of FP$5$ and changes in funding rules of FP$6$. It is verified in fig. \ref{fig:fig3b} as well, where there is a decrease in the number of triangles in the real data around that date. It is also worth noting that the patent triangles are practically zero in the real data and few even in the shuffled ones. This is due to the very high number of triangles in the FP layer, which means that practically any inter-regional collaboration link in the patent layer created will very likely have a corresponding one in the FP layer. Thus, and owing to the approach used, almost all patent links are transformed to common ones and only few remain with no corresponding ones. 

\begin{figure}[!ht]
	\begin{center}
		\subfloat[\label{fig:fig4a}]{{\includegraphics[width=0.49\textwidth]{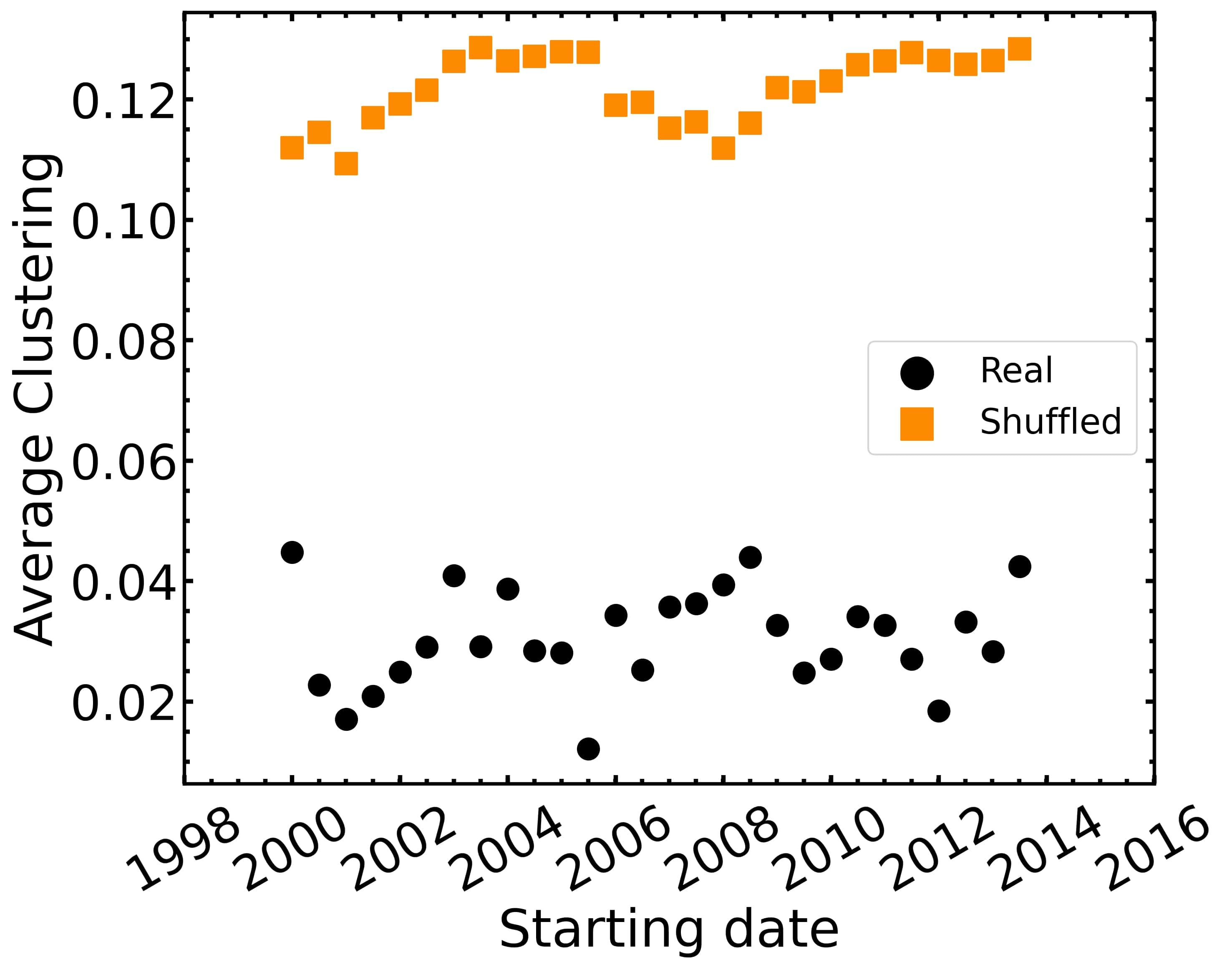} }}
		\subfloat[\label{fig:fig4b}]{{\includegraphics[width=0.49\textwidth]{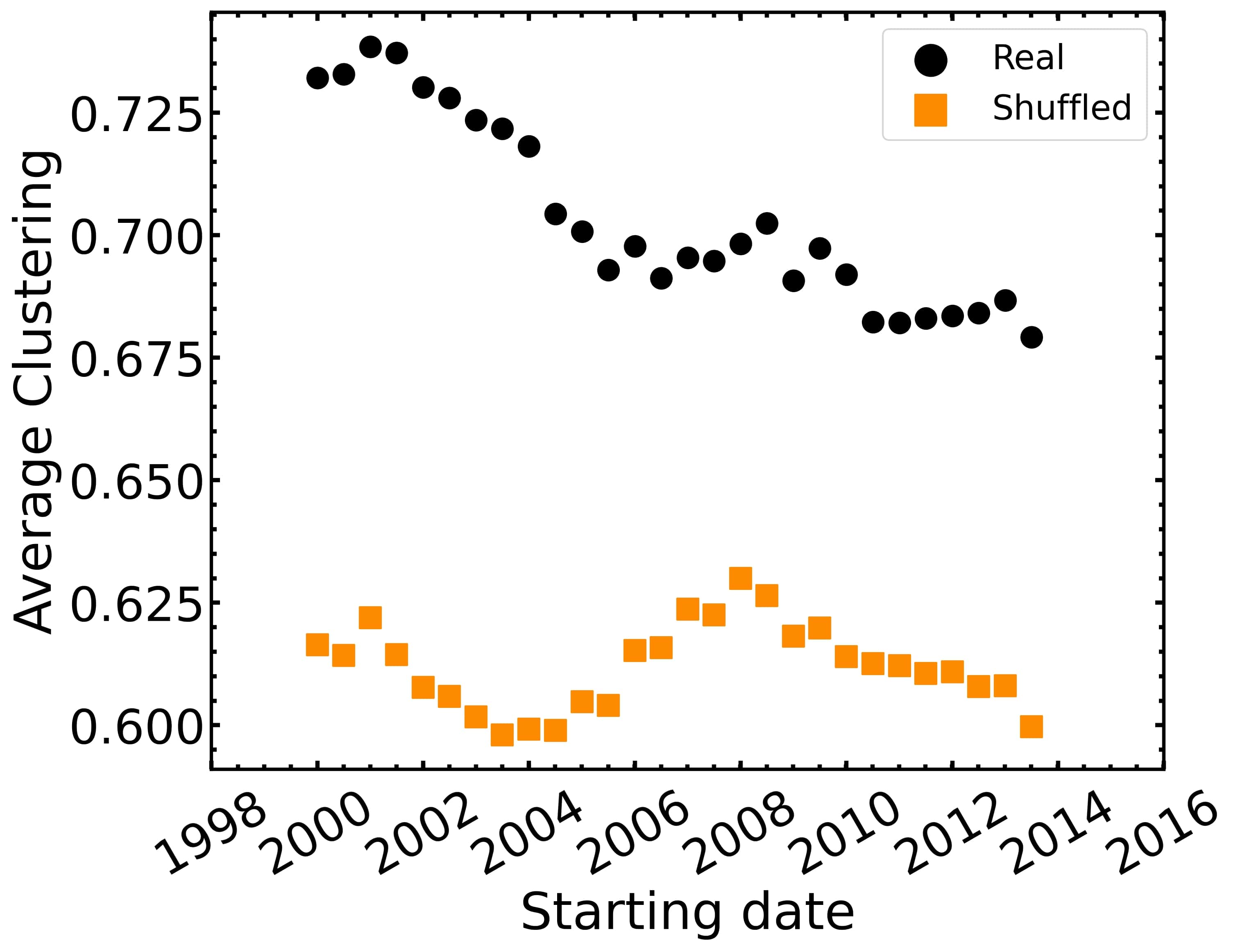} }} \\
		\subfloat[\label{fig:fig4c}]{{\includegraphics[width=0.49\textwidth]{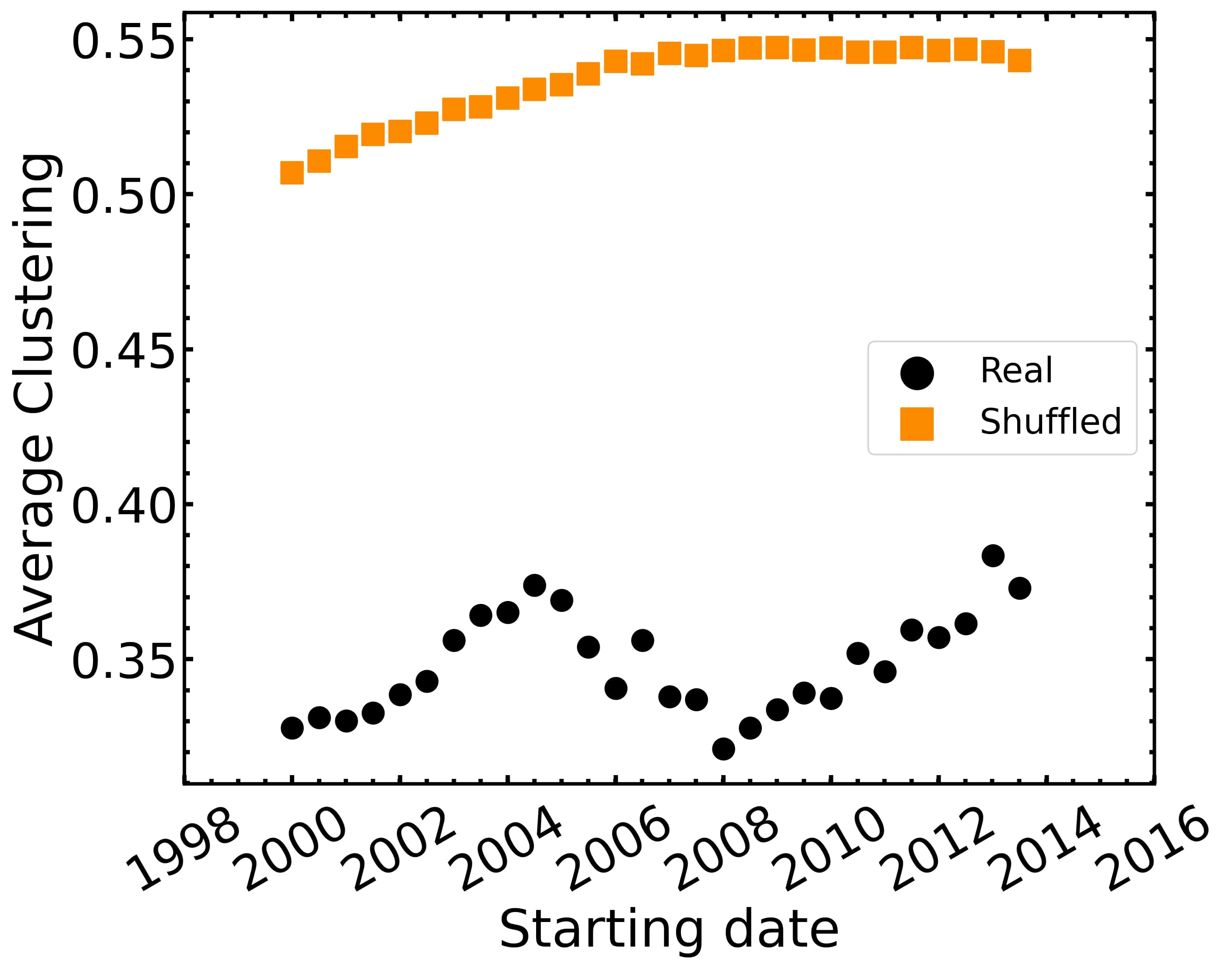} }}
		\subfloat[\label{fig:fig4d}]{{\includegraphics[width=0.49\textwidth]{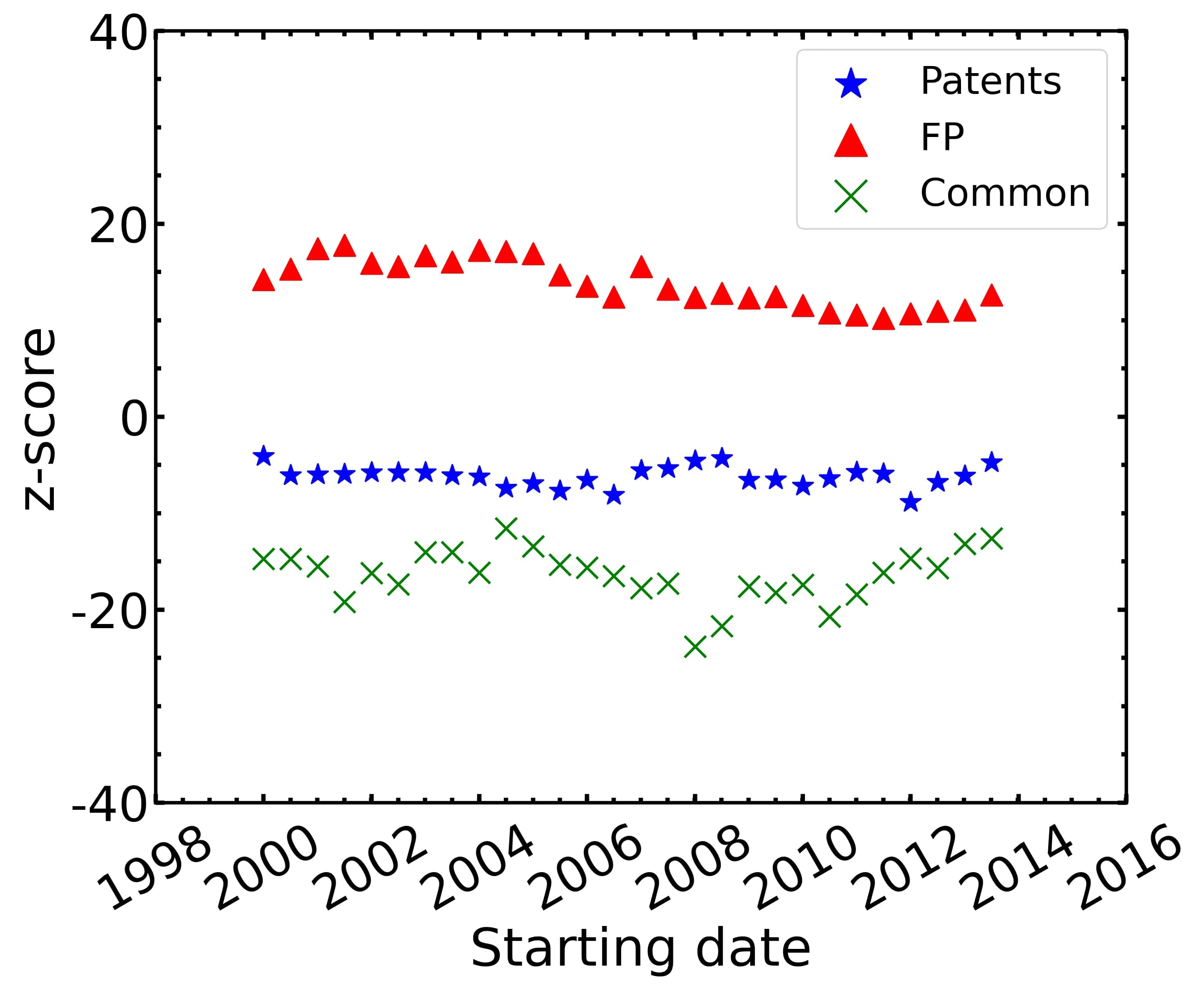} }}
		\caption{The average local clustering coefficient calculated at the end of the $28$ windows versus their starting dates for a) patents, b) FP, and c) their common network. Black dots represent the real data and orange squares represent the average of $50$ shuffled networks. 
			d) z-score of the average local clustering coefficient for the patent (stars), FP (triangles), and Common (cross) network versus time.
		}
		\label{fig:fig4}
	\end{center}
\end{figure}

Next, we repeat the same process by calculating each node's local clustering coefficient of each layer, including their common network. This multiplex network is again formed by removing the links that exist in both the patent and the FP layers and inserting them into the common one. We then calculate the average value of the local clustering coefficients of each layer and of the common network, fig. \ref{fig:fig4}. Real data in patents, fig. \ref{fig:fig4a}, and the common network, fig. \ref{fig:fig4c}, show that they clearly have smaller averaged local clustering coefficients and, thus, when compared to shuffled data, do not tend to form triangles. As for the FP layer, fig. \ref{fig:fig4b}, we notice that real data exhibit a higher averaged local clustering coefficient, as compared to the shuffled data. The z-score of the local clustering coefficient, fig. \ref{fig:fig4d}, clears up the question whether such results could be randomly obtained. The values shown point to a non randomized process for all three cases, and the existence of an underlying preferential type of mechanism for the growth of these systems. 

Fig. \ref{fig:fig4}, when compared with fig. \ref{fig:fig3}, shows some similarities in a qualitative sense, as both show the same type of preference in real over shuffled triangle formation for FPs and the opposite for patents and the common network. However, the quantitative difference is significant as the triangles approach shows much more strongly the existence of a non random process. It shows a $20-50\%$ preference in real over shuffled FP triangles, while real $\overline{C_i}$ values are only $10-20\%$ higher than their respective shuffled ones. Similarly, for the patents and the common network the number of triangles is about $10$ and $3-4$ times higher in the shuffled data than in the real ones, while $\overline{C_i}$ values are only $3-4$ and less than $2$ times more, respectively. z-score values are similar in both figures with the triangle ones, fig. \ref{fig:fig3d}, being again larger than the averaged local clustering coefficient ones, fig. \ref{fig:fig4d}.

\begin{figure}[!ht]
	\begin{center}
		\subfloat[\label{fig:fig5a}]{{\includegraphics[width=0.33\textwidth]{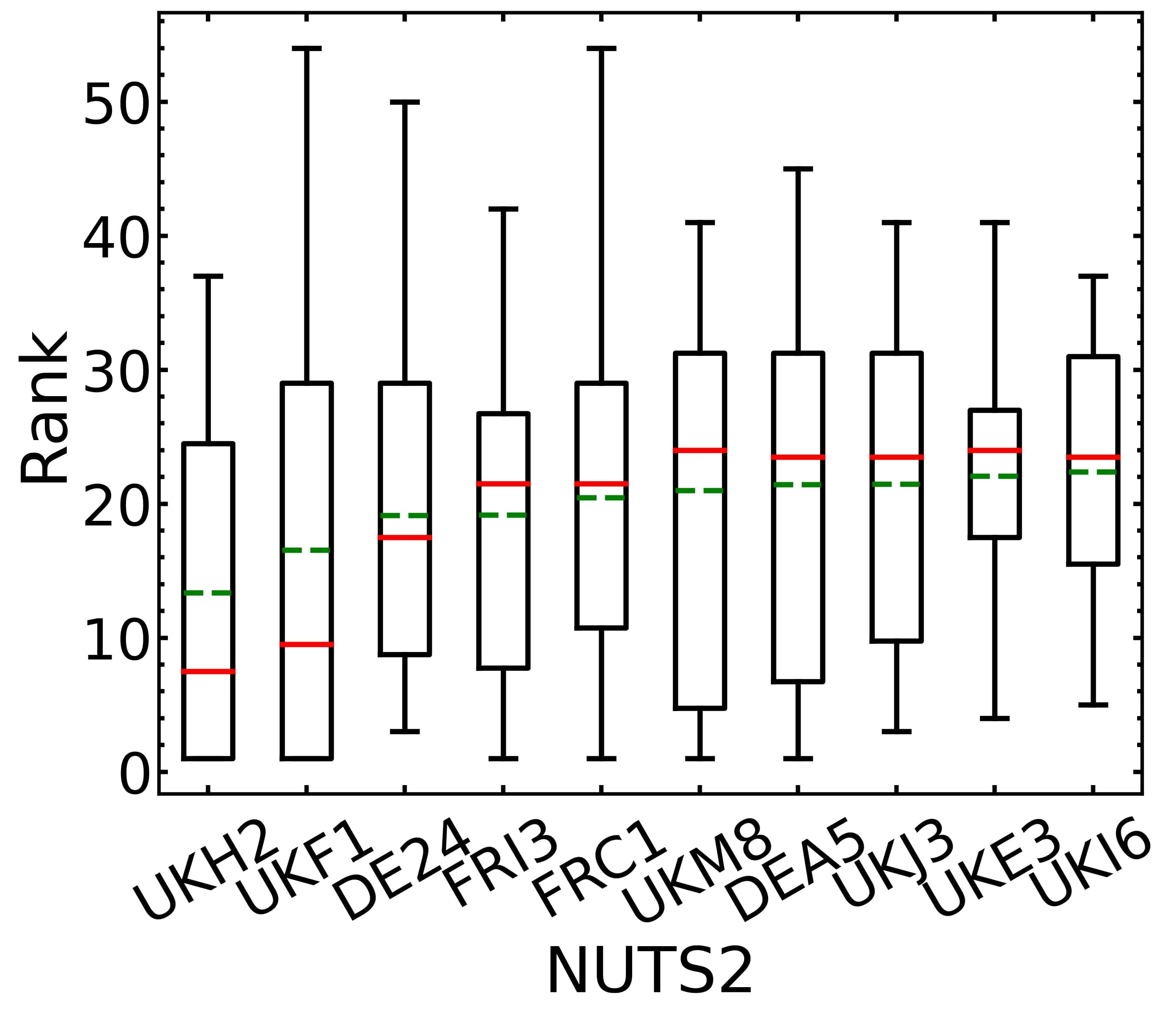} }}
		\subfloat[\label{fig:fig5b}]{{\includegraphics[width=0.33\textwidth]{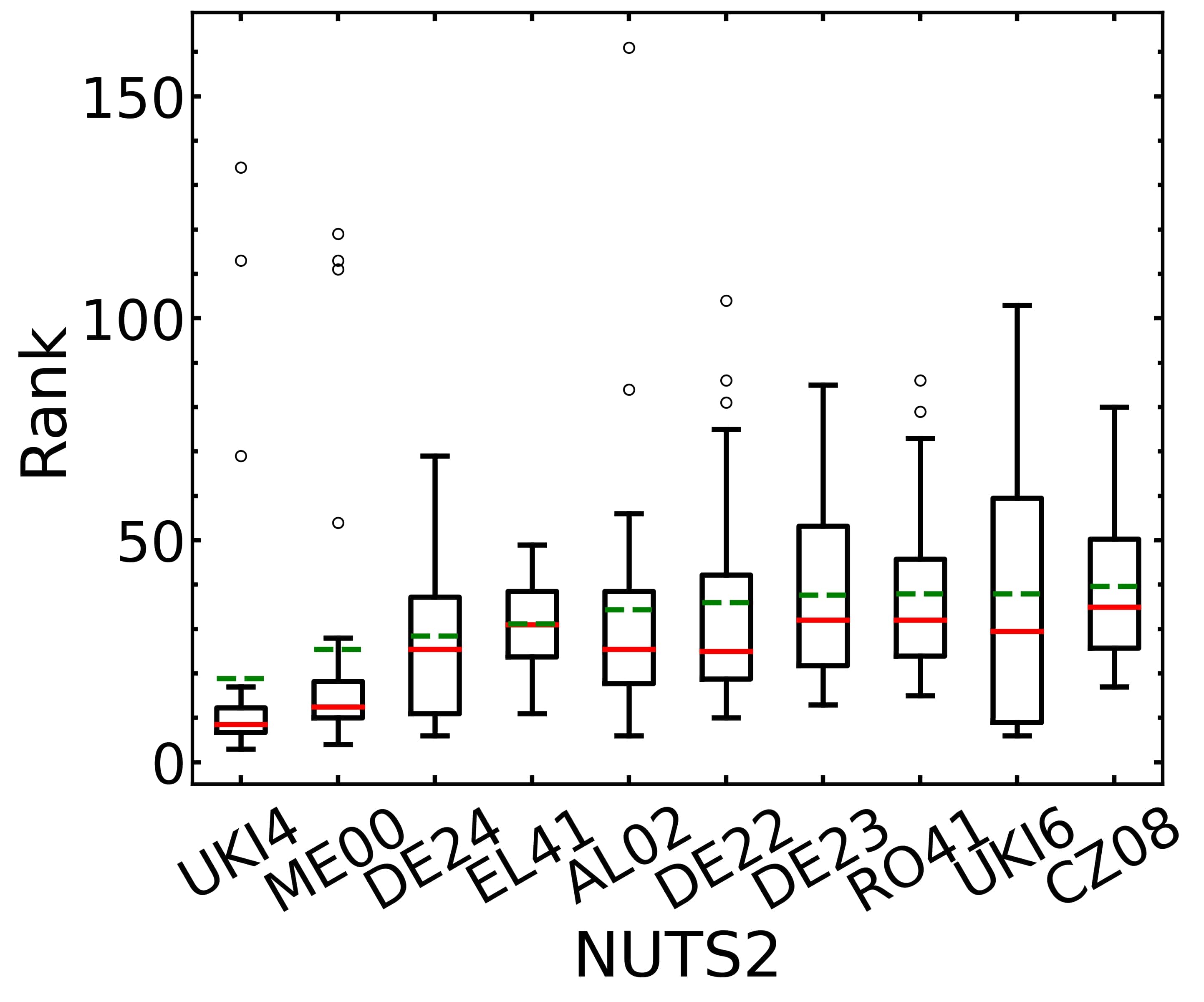} }} 
		\subfloat[\label{fig:fig5c}]{{\includegraphics[width=0.33\textwidth]{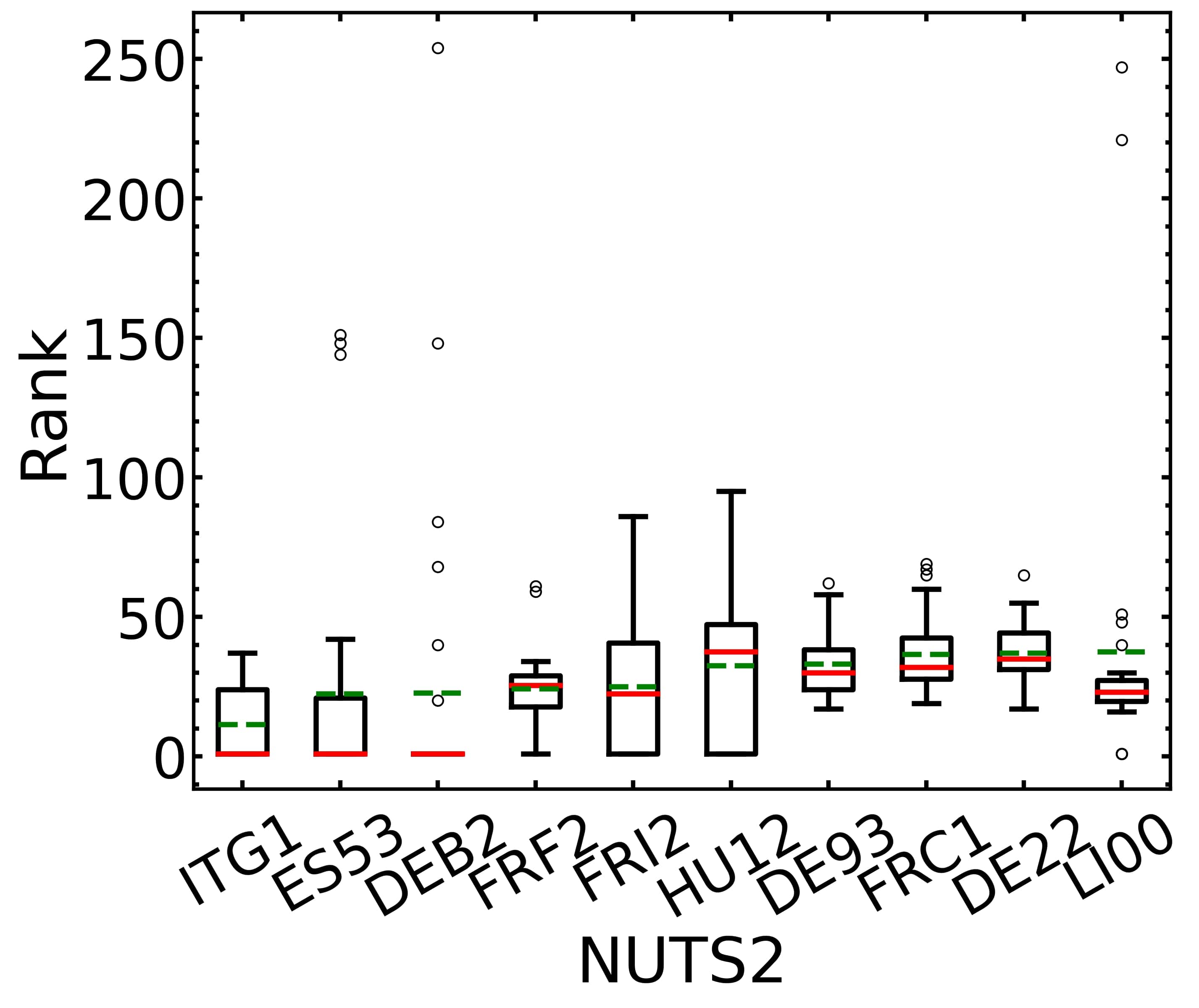} }}
		\caption{ 
			Boxplot of the top $10$ NUTS$2$ regions for a) patents, b) FPs, and c) their common network in terms of average ranking, according to the local clustering coefficient ranking of the $28$ windows (green dashed lines). Red lines show the median value of their ranking for the $28$ windows.}
		\label{fig:fig5}
	\end{center}
\end{figure}
\begin{figure}[!ht]
	\begin{center}
		\subfloat[\label{fig:fig6a}]{{\includegraphics[width=0.33\textwidth]{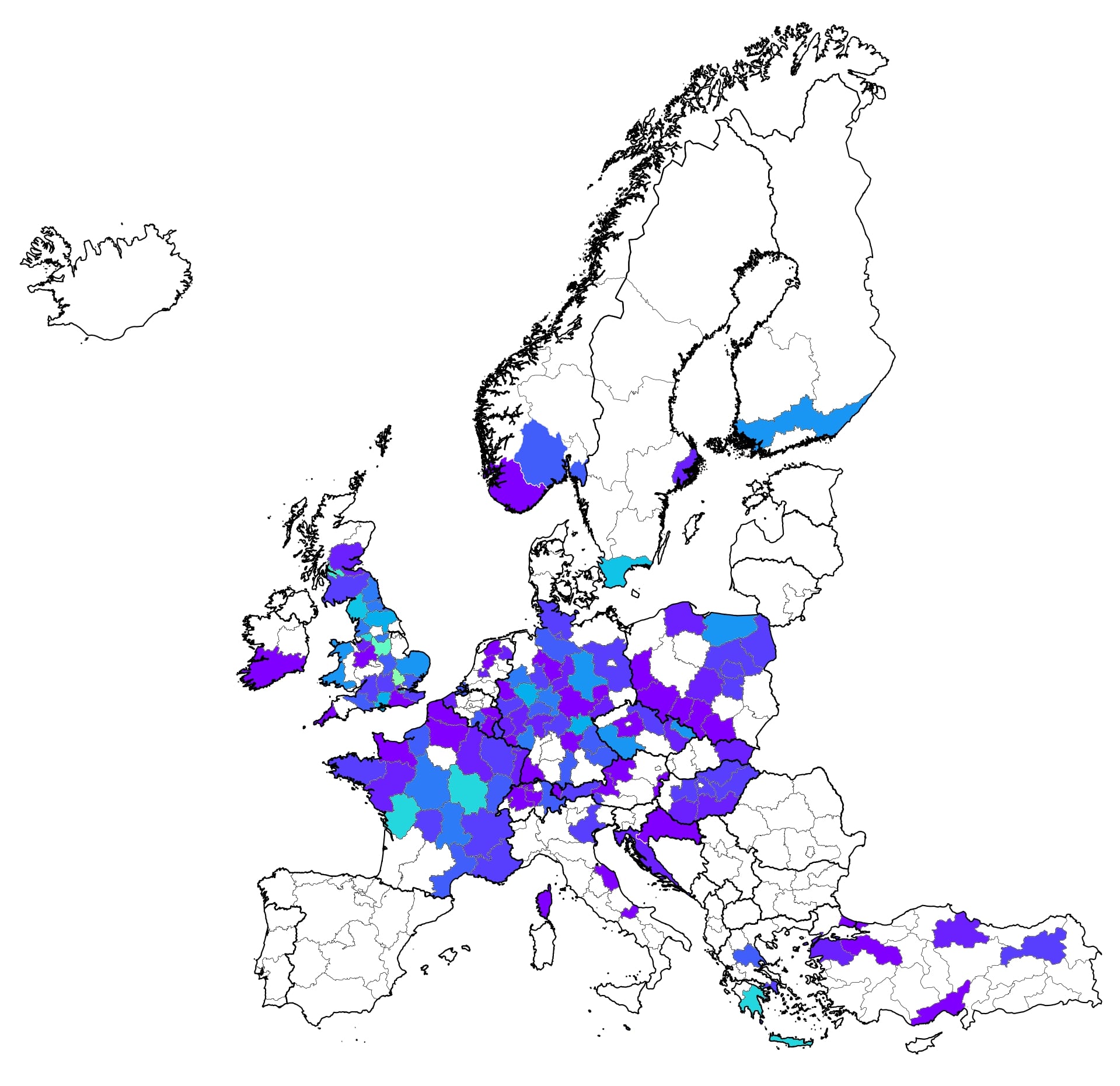} }}
		\subfloat[\label{fig:fig6b}]{{\includegraphics[width=0.33\textwidth]{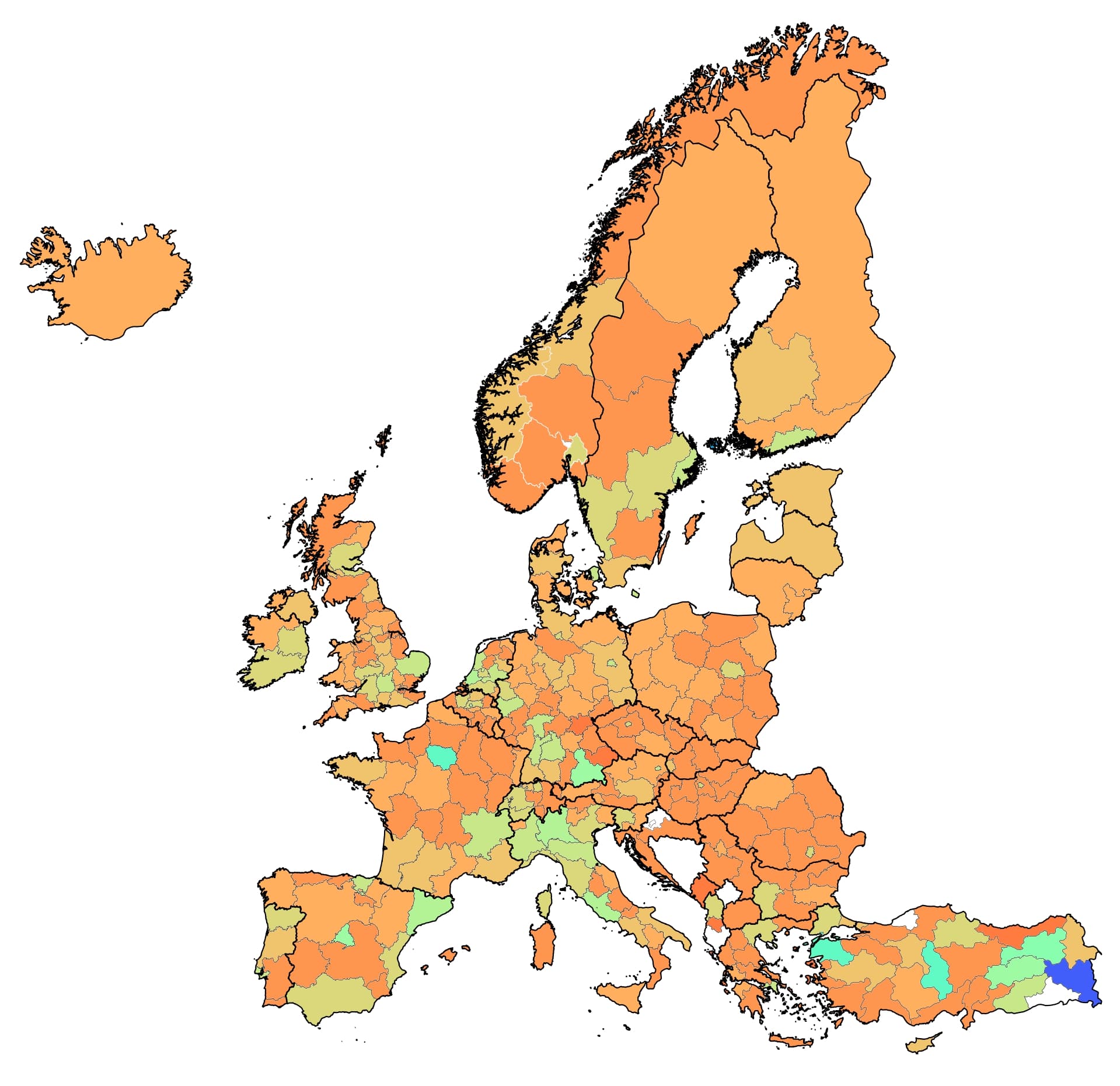} }} 
		\subfloat[\label{fig:fig6c}]{{\includegraphics[width=0.33\textwidth]{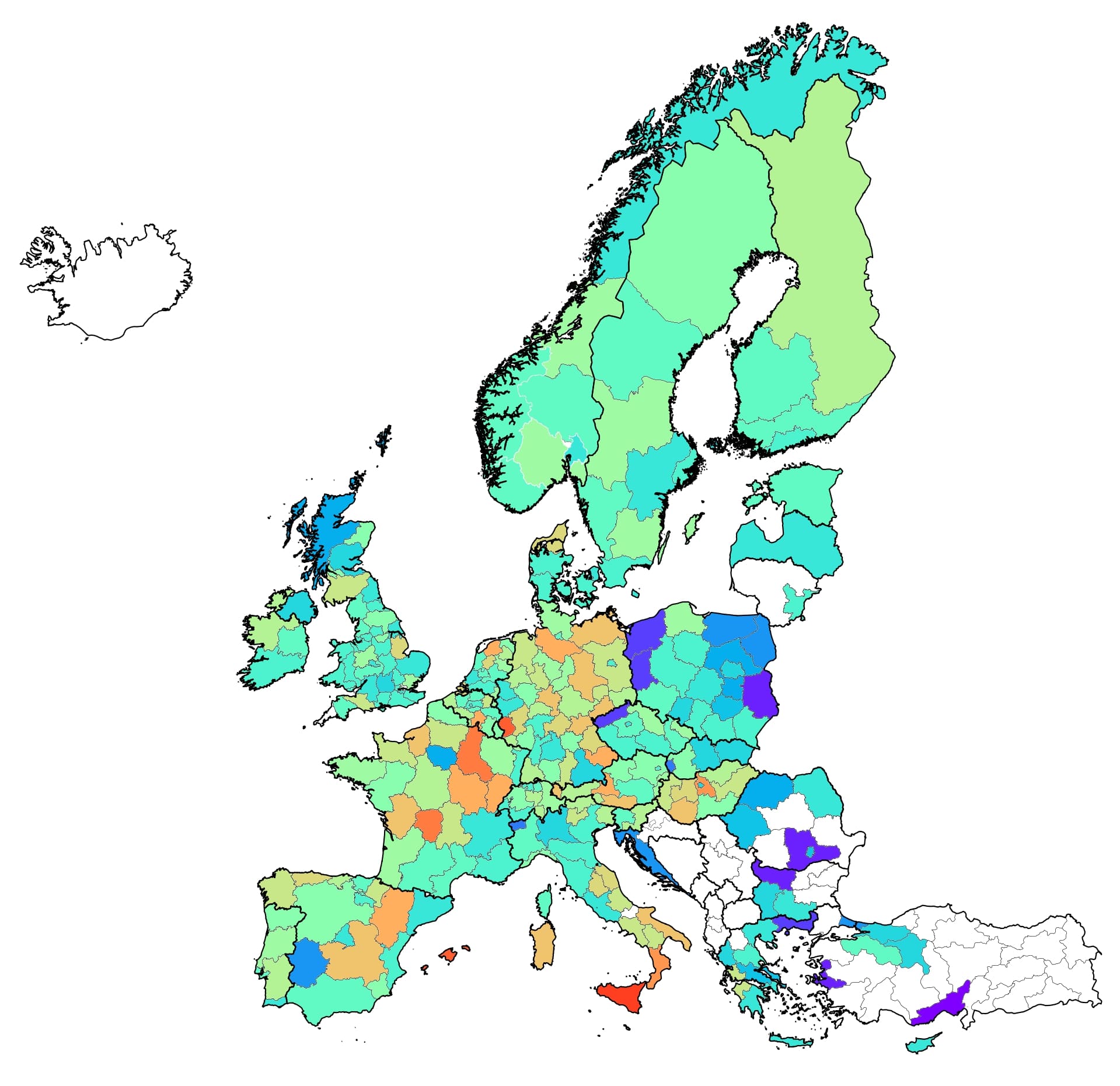} }} \\
		\subfloat[\label{fig:fig6d}]{{\includegraphics[width=1\textwidth]{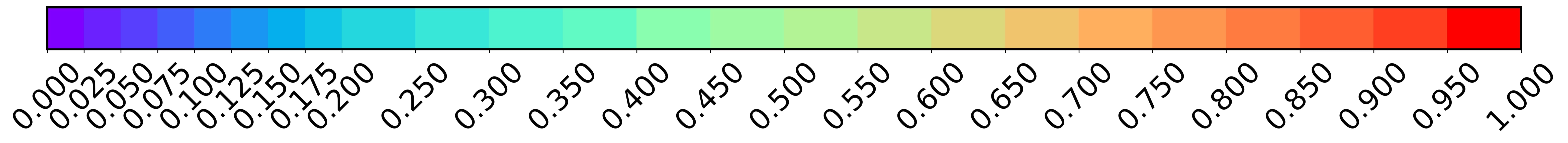} }}
		\caption{(Color online) European color-based map of the averaged local clustering coefficient of each NUTS$2$ region. Real data for a) patents, b) FPs, and c) their common layer. As the average value of the local clustering coefficient increases, the colors change according to the d) colorbar (from left to right). White color indicates regions with a zero value in the clustering coefficient, which corresponds to zero triangles in that layer after the removal of the common triangles, or zero common triangles.}
		\label{fig:fig6}
	\end{center}
\end{figure}

Our next goal is to identify at each layer the nodes with the higher average rank ($\textrm{rank}_{aver}$), according to their local clustering coefficient rank for the $28$ windows ($\textrm{rank}_{sb}$) This will help us pinpoint the key nodes in triangular collaborations over time. More specifically, for each window we rank the nodes according to their local clustering coefficient, $\textrm{rank}_{sb}$. We allow for two, or more, nodes to have the same rank if needed. We then average over all the $28$ ranking positions of each node in each sub-network, and then re-rank the averaged data, $\textrm{rank}_{aver}$. Figure \ref{fig:fig5} shows the boxplots of the top $10$ NUTS$2$ regions for all $3$ cases (patents, FPs, common network), which are sorted from left (higher $\textrm{rank}_{aver}$) to right (lower $\textrm{rank}_{aver}$) according to the average value of their rank. We notice that there are very few NUTS$2$ regions that exist in all $3$ layers.\\
Finally, we present $3$ European maps, one for each case, colored according to the averaged (out of the $28$ windows) local clustering coefficient of each NUTS$2$ region, fig. \ref{fig:fig6}. We notice that regions with intense scientific history, as for example Paris, are not among the highest regions in the FPs and the common network. Although this may not seem reasonable, it is due to the fact that we examine the existence of triangles (through the local clustering coefficient), and not that of links which may actually be too many (in this case in Paris there are 324 links, which ranks Paris as the first region in number of links). We also notice that there are regions, or even entire countries, that may be part of triangular type of scientific collaborations in the FPs at a much higher than expected rate. However, their part in the actual number of patents is relatively small as compared to that of other regions (Iceland, and most of Turkey).\\
In the supplementary material we also show the same figure, fig. S3, produced with a different goal in mind. Specifically, we include the common triangles in the patent and FP layers rather than remove them, in order to have a clearer view of these two layers only, each on their own, and not the common one. The results show that the patent layer is almost identical to the results of the common network, fig. \ref{fig:fig6c}, while the FP layer shows even higher local clustering coefficient values for most regions. \\
The supplementary material also presents the results for the top-ten NUTS$2$ regions ranked in order of participation in collaboration triangles for FPs, patents and their common network in fig. S4. The average number of triangles, shown in fig. S5, is normalized over the maximum value of the average number of triangles, while the averaging is done over the same $28$ windows as before. The results paint a slightly different picture than that of the local clustering coefficient. They emphasize on the role of major urbanized city centers found on all EU countries which simply gather much larger numbers in links belonging to triangles in the FP layer and the common network. One can notice the major differences between regions such as Sicily, Paris, Rome, Madrid, and many more in fig. \ref{fig:fig6} as opposed to fig. S5. Such a representation of collaborations is closer to those of studies doing a simple statistical analysis of regions in either patents or subsidized research.

\section{Conclusions}\label{conclusions}
In summary, we study the evolution of triangles in a multiplex network consisting of patents and European Framework Programmes to uncover any preference in forming extended collaborations of triangular form, rather than dual ones of simple link type. In addition, we compare the results to those of the local clustering coefficient in order to identify any differences.

The results show that in the patent layer, the number of triangles formed only on that specific layer is extremely small in comparison to those of the FP layer and the common network. In addition, when comparing the real data to its shuffled versions, one concludes that in such a real system there is a strong preference to form isolated links rather than entire triangles. In fact, the number of triangles in the shuffled data is many times larger (up to $10$ times). Similar behavior is noticed in the common network, where the shuffled data form significantly more triangles than the real ones (up to $4$ times). On the contrary, FP collaborations show a stronger preference to form triangles in the real, rather than the shuffled, layers. We also study the local clustering coefficient, which yields qualitatively quite similar results, yet in all cases with much less distinguishable differences. Indeed, by studying the number of triangles one can emphasize on the differences between real and shuffled networks when compared to their local clustering coefficient counterparts. The main point observed in the system of research and innovation studied here is that triangles elucidate more easily than the clustering coefficient the preference (or avoidance) in triangular forms of collaboration over simple dual ones.

We want to identify the NUTS$2$ regions that have constantly, for all time windows, proven to have a high local clustering coefficient, and can, thus, be considered significant for the creation of their networks. What we see is that for each layer the top $10$ regions are mostly different, and there are very few which are common in all three layers. Furthermore, according to the averaged value of the local clustering coefficient from all the $28$ windows, we notice some regions with counter-intuitive behavior. There are regions of high scientific activity with low local clustering coefficient, and other regions that are not so scientifically active which have a higher local clustering coefficient value. The same result can be seen clearly by comparing fig. \ref{fig:fig6} with fig. S5. This could be due to the fact that the latter regions may have a very small number of collaborations, but possibly the same exact regions over time, thus forming relatively high numbers of triangles. On the contrary, scientifically active regions may have many more collaborations, but these do not always form enough triangles, resulting on lower local clustering coefficient values.

Our research results and our methodology can help funding authorities and policy makers decide on whether specific regional actions need to be taken to support specific geographical areas. By adding a new criterion that can in some cases identify differences in real versus randomized networks much easier, the results of this study can prove valuable even to other systems. For example, in cases of real social multiplex networks, it can perhaps help identify in a new way whether there is a preference or not for the friend of a friend to be a friend. In any case, since the results hold true for both a sparse (patent) and a dense (FP) layer, as well as their common multiplex network, it is possible that many other such systems can use our results and methodology.


\section{Acknowledgments}\label{acknowledgments}
Results presented in this work have been produced using the Aristotle University of Thessaloniki (AUTh) High Performance Computing Infrastructure and Resources.

\end{document}



\begin{center}
    \textbf{\huge Supplementary Material}
\end{center}

\begin{figure}[H]	
	\begin{center}
		\subfloat[\label{fig:fig1a}]{{\includegraphics[width=0.32\textwidth]{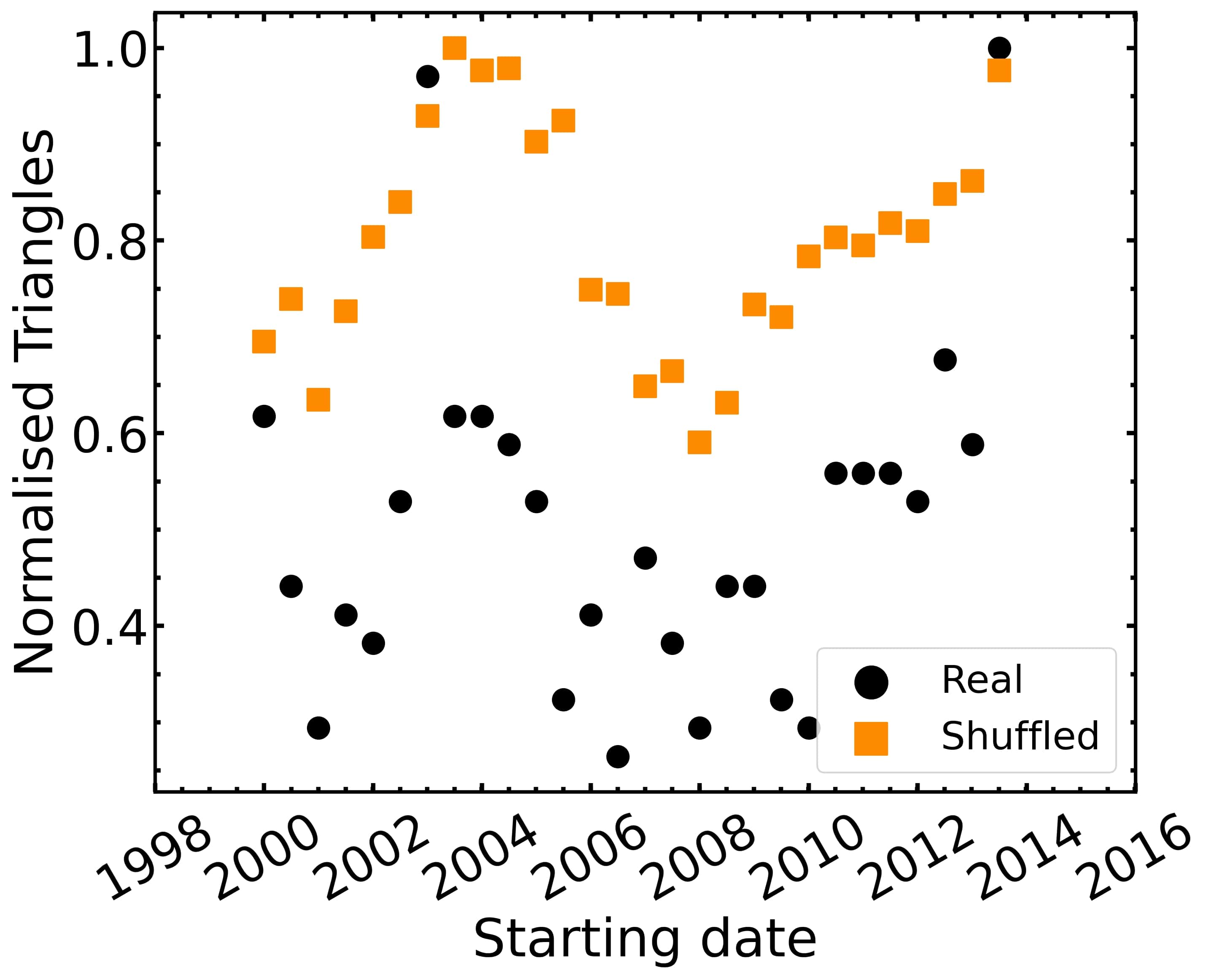} }}
		\subfloat[\label{fig:fig1b}]{{\includegraphics[width=0.32\textwidth]{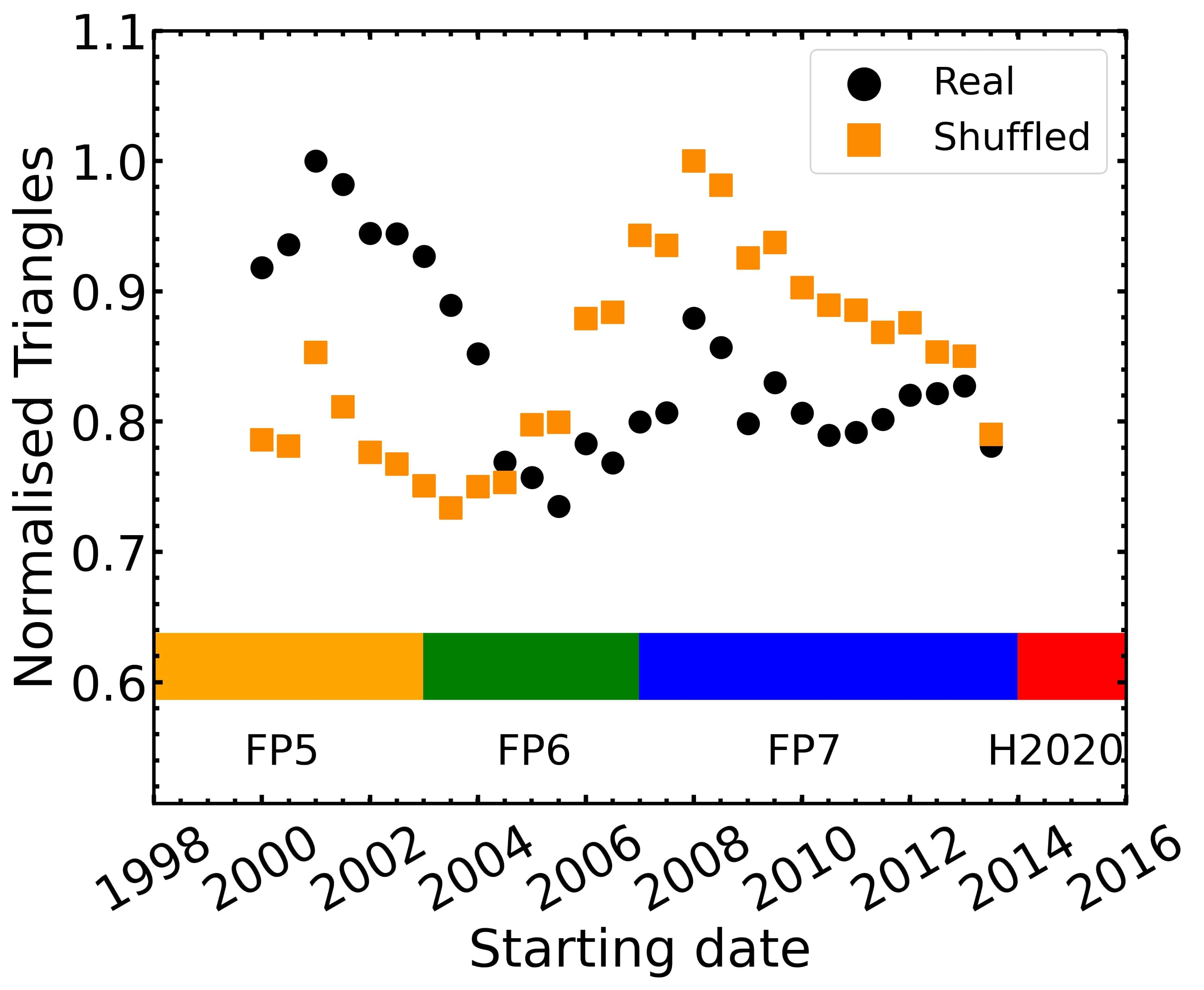} }} 
		\subfloat[\label{fig:fig1c}]{{\includegraphics[width=0.32\textwidth]{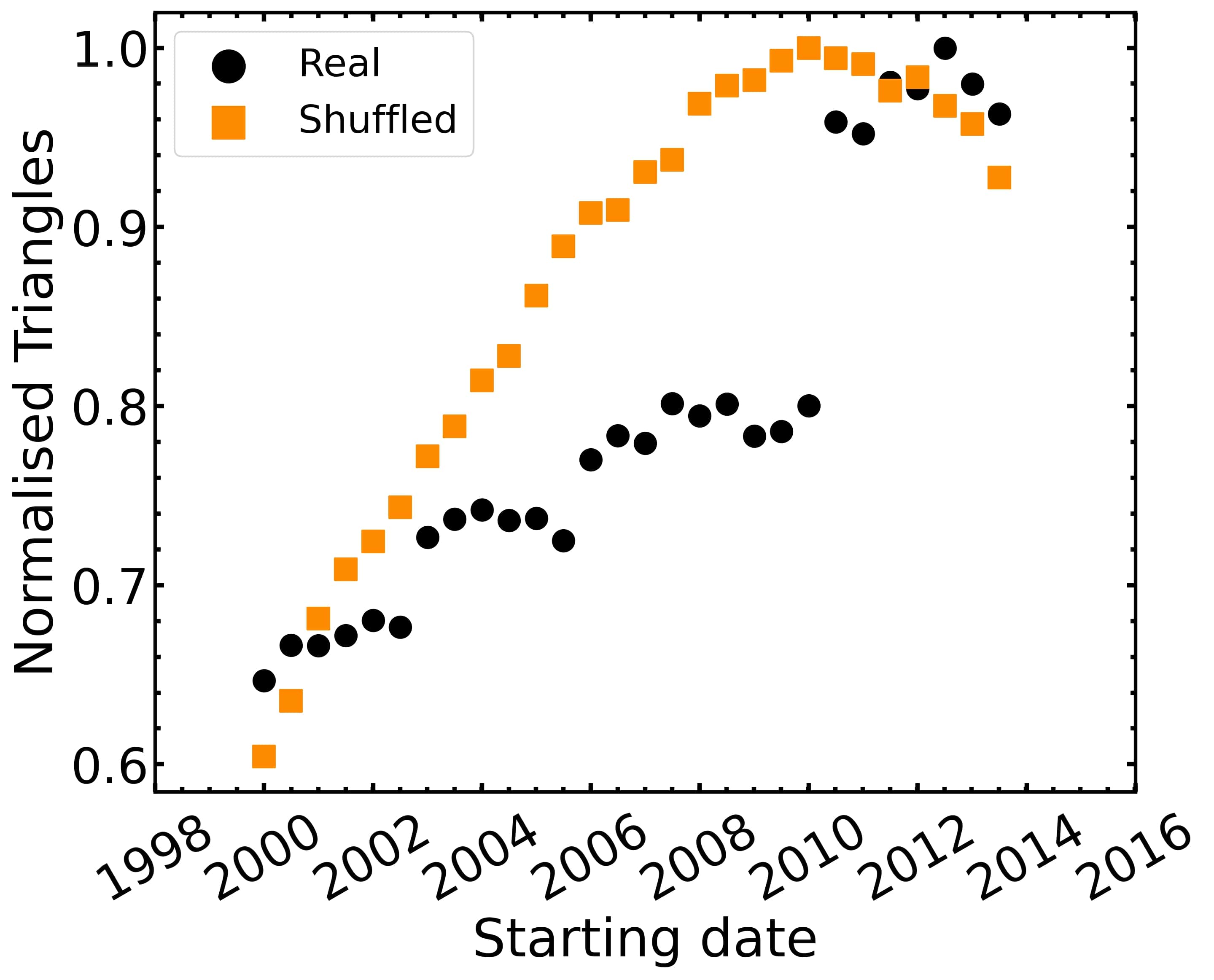} }}
		\caption{The normalized number of triangles at the end of the $28$ windows, versus their starting dates for a) patents, b) FP, and c) their common network. Black dots represent the real data, and orange squares represent the average of $50$ shuffled networks. The rectangular colored areas (colors are random) at the bottom of figure b show the FPs duration of each.}
		\label{fig:fig1}
	\end{center}  	
\end{figure}

In the supplementary material, we provide some graphs that help explain the results of the main paper, as well as study the same system with minor only changes. Figure \ref{fig:fig1} shows the number of triangles at the end of each of the $28$ sub-networks normalized over their respective maximum values. This result, in contrast to the actual number of triangles, which is presented in figure $2$ of the main manuscript, helps understand the relative trend of their evolution. In fact, it shows that while for the real data of patents the number of triangles exhibits much lower values than the shuffled ones, the evolution of both results over time is quite similar. For the FP results, the ratio of real data triangles over shuffled ones is greatly reduced. While up to $2004$ the real data are relatively more than the shuffled ones, this changes near $2005$. This is the period where late FP$6$ and early FP$7$ projects start to dominate and shape the FP layer (due to an average $3.5$ year duration of these projects). The dynamics of the real vs shuffled change again during the end of our study period. This may be due to Horizon funding scheme changes being implemented and counted in the layer. As for the results of the common network, they seem to be consistent with $4$ periods throughout which the number of triangles is constant. It also seems there is an abrupt change from one to another, and that a similarity exists between the duration of each such period and the FP duration, with a small added temporal delay of $2-3$ years.

\begin{figure}[H]
	\begin{center}
		\subfloat[\label{fig:fig2a}]{{\includegraphics[width=0.49\textwidth]{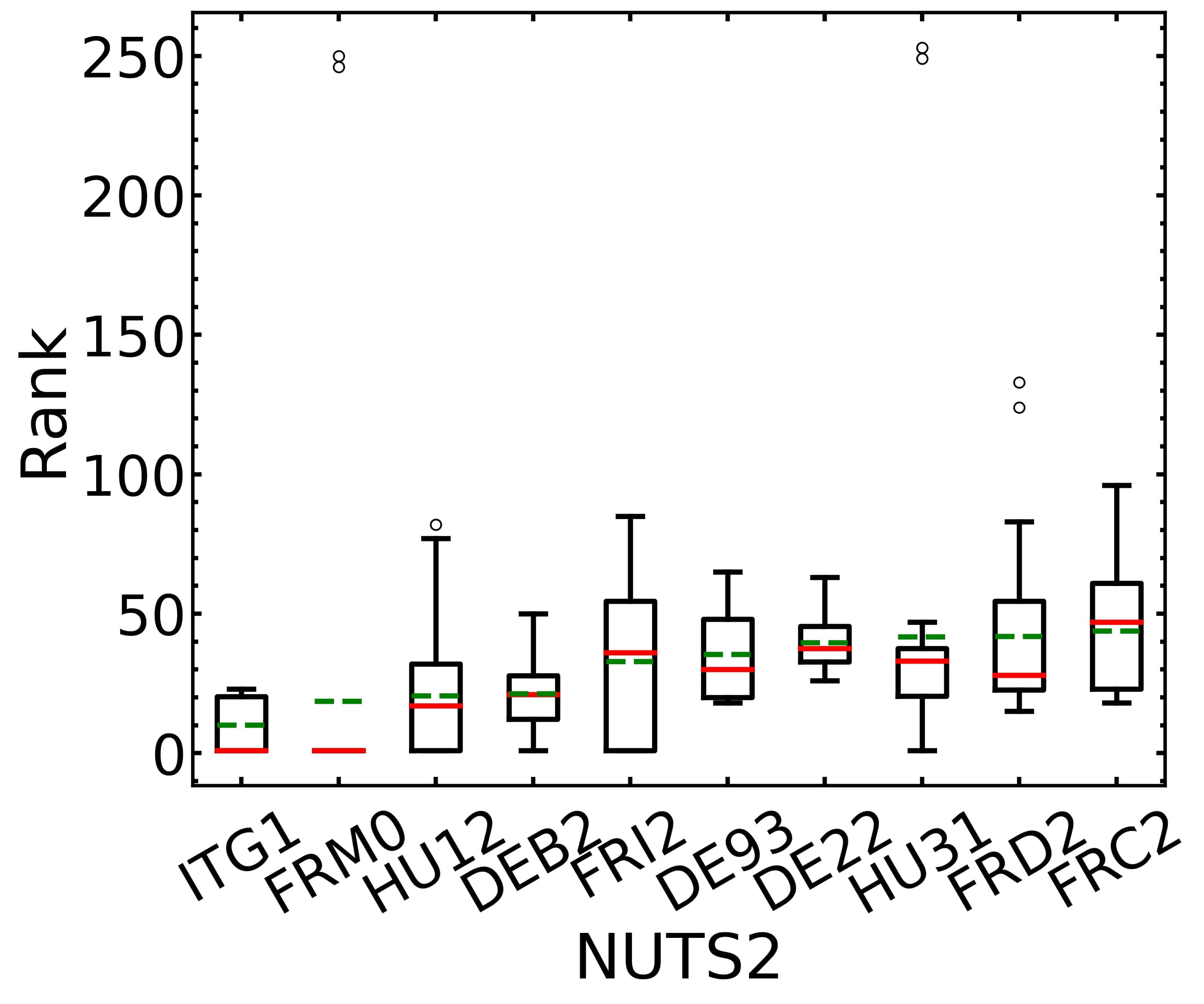} }}
		\subfloat[\label{fig:fig2b}]{{\includegraphics[width=0.49\textwidth]{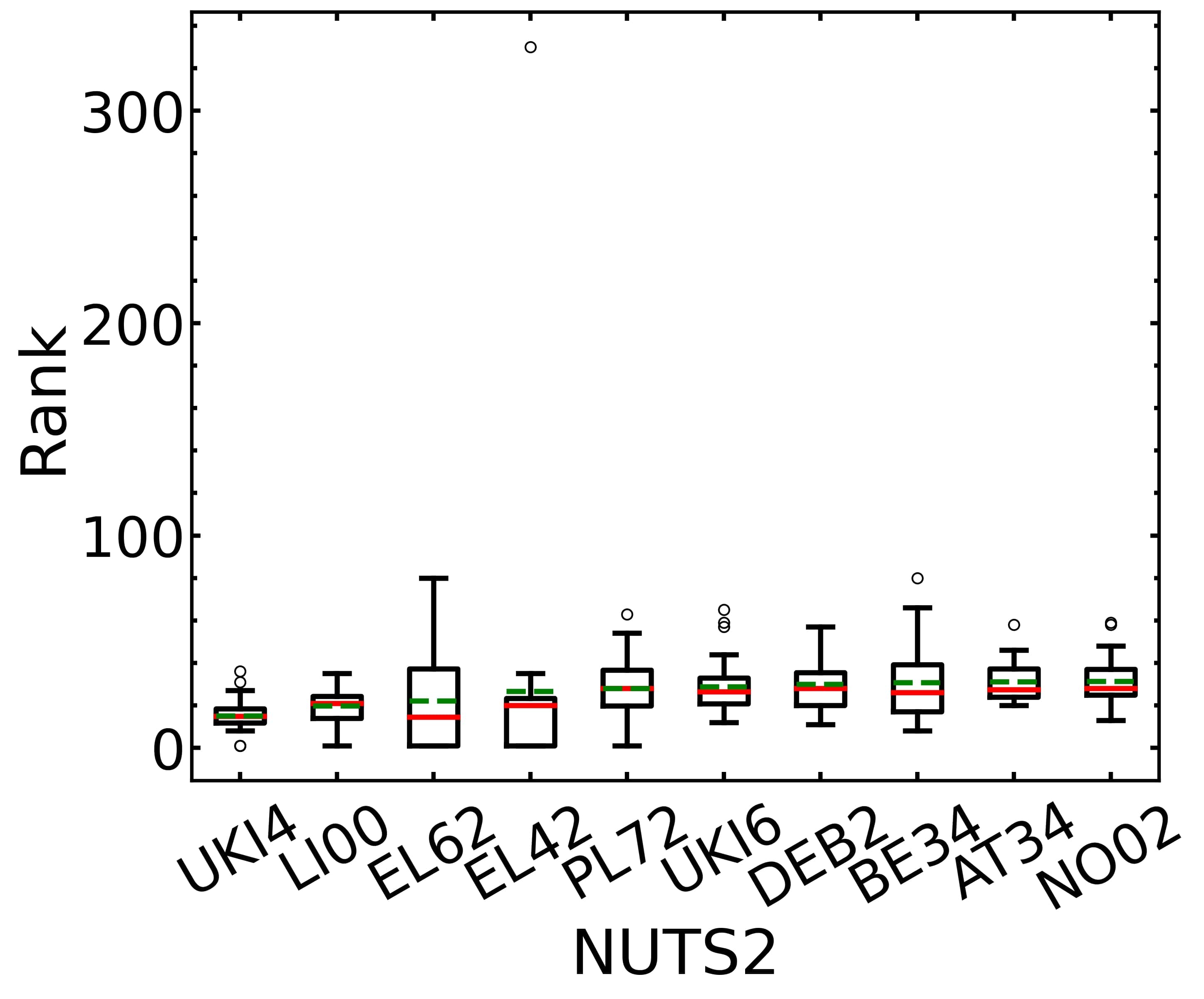} }} \\
		\caption{Boxplot of the top $10$ NUTS$2$ regions for a) patents, and b) FPs in terms of average ranking, according to the local clustering coefficient ranking of the $28$ windows (green dashed lines). Red lines show the median value of their ranking for the $28$ windows.}
		\label{fig:fig2}
	\end{center}
\end{figure}

\begin{figure}[H]
	\begin{center}
		\subfloat[\label{fig:fig3a}]{{\includegraphics[width=0.5\textwidth]{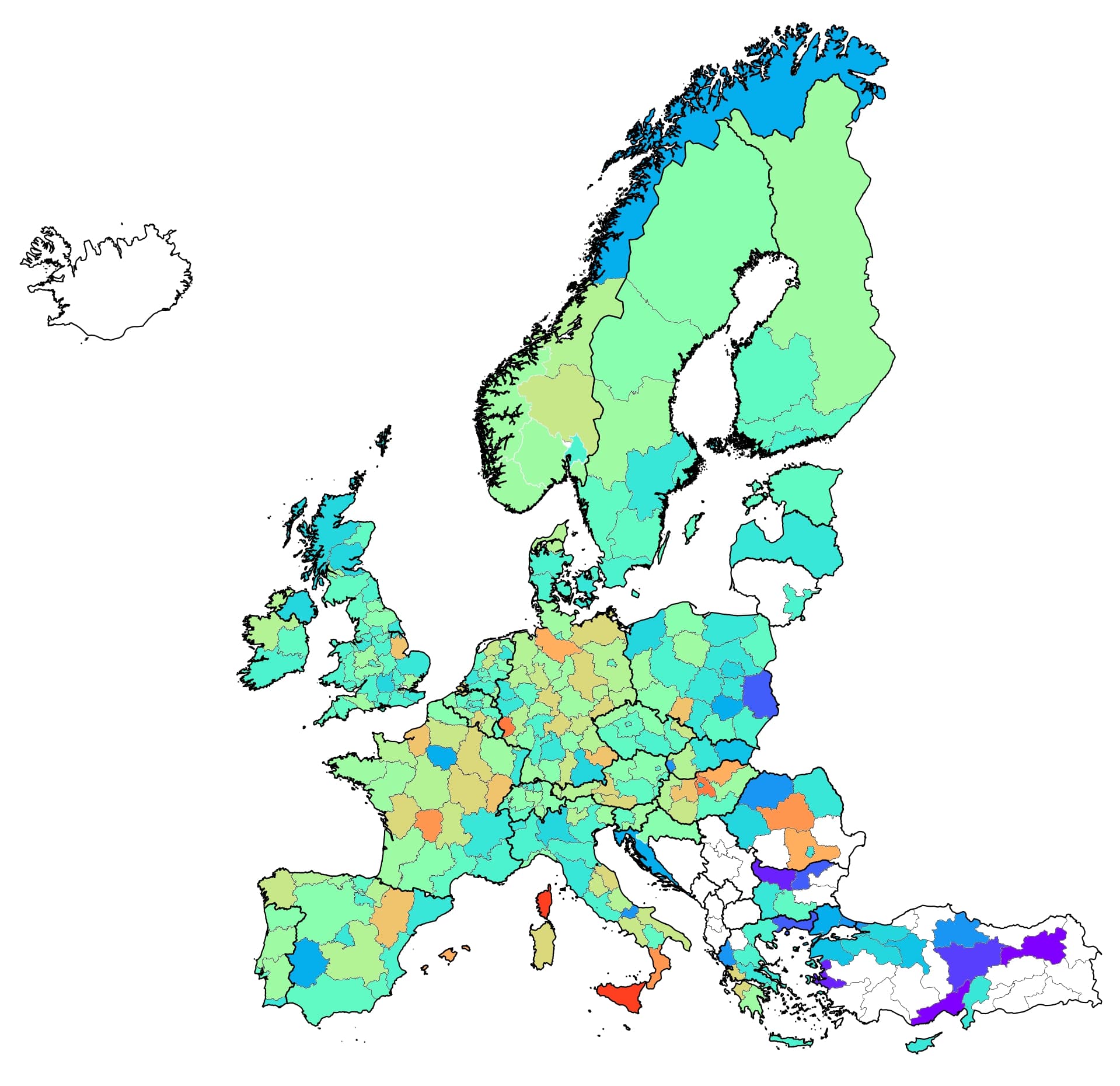} }}
		\subfloat[\label{fig:fig3b}]{{\includegraphics[width=0.5\textwidth]{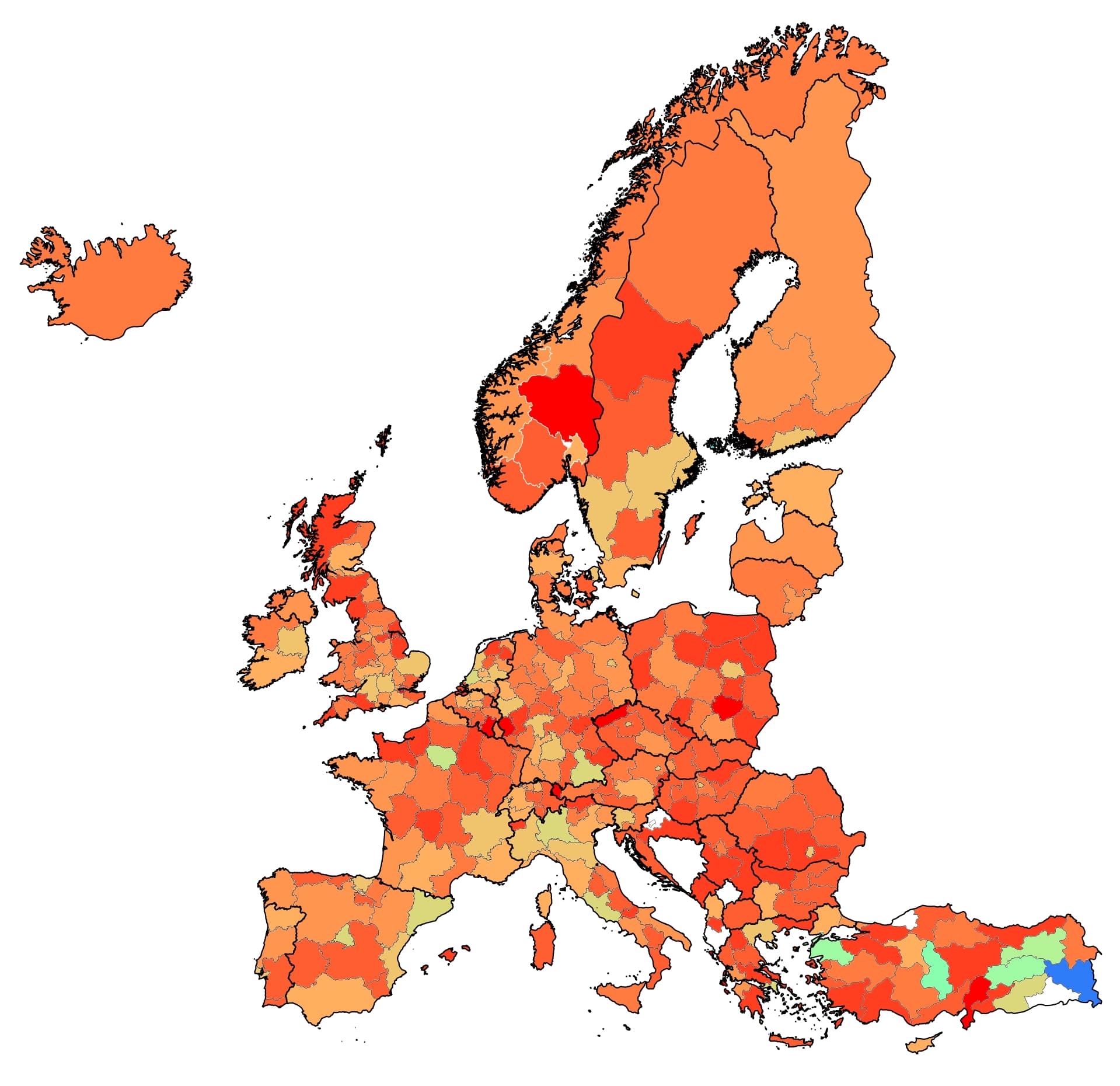} }} \\
		\subfloat[\label{fig:fig3c}]{{\includegraphics[width=1\textwidth]{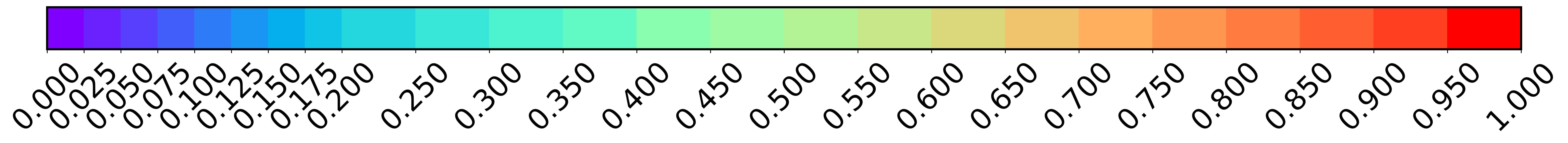} }}
		\caption{(Color online) European colour-based map of the averaged local clustering coefficient of each NUTS$2$ region. Real data for a) patents, and b) FPs. As the average value of the local clustering coefficient increases, the colors change according to the d) colorbar (from left to right). White color indicates regions with a zero value in the clustering coefficient, which corresponds to zero triangles in that layer after the removal of the common triangles.}
		\label{fig:fig3}
	\end{center}
\end{figure}

Next, we present two figures similar to those of figures $4$ and $5$ of the main manuscript. Their difference is that in the main text, the rank of each NUTS$2$ region is being calculated when the common links between the FP and the patent layer have been removed. Here, we show what these figures would be like if we studied the two layers separately (no links extracted from either layer). Figure \ref{fig:fig2} shows the top $10$ NUTS$2$ regions according to their averaged local clustering coefficient value ranking, over all $28$ windows. Figure \ref{fig:fig3}, shows what the average local clustering coefficient would be for each NUTS$2$ region, if we had not removed the common links. 

It is worth noting that the top $10$ ranked NUTS$2$ regions in terms of local clustering coefficient values are different in the main text than they are here. In fact most of the top $10$ regions of the main texts' common network are now located in the FP and patent top $10$. Indeed, the patent and FP rankings have significantly changed. This is expected for the patent layer, and is due to the much higher number of triangles in the common network than on the patent layer of the main text. It is not, however, straightforward for the FP network which already has a very high number of triangles and relatively few are not extracted now (the number of triangles in FP layer reaches $1.6$ million and the common network up to $40,000$ only). This is due to the fact that the common network, as analyzed in the main text, has extracted links responsible for higher local clustering coefficient values in the FP layer. 

The overall results of the recalculated local clustering coefficient show that the patent European map of the supplementary material (\ref{fig:fig2a}) greatly resembles the common network map of the main text (fig. $6c$). As for the FP European map, it shows here a significantly higher value of local clustering coefficient than it does in the main text (fig. $5b$).

\begin{figure}[H]
	\begin{center}
		\subfloat[\label{fig:fig4a}]{{\includegraphics[width=0.45\textwidth]{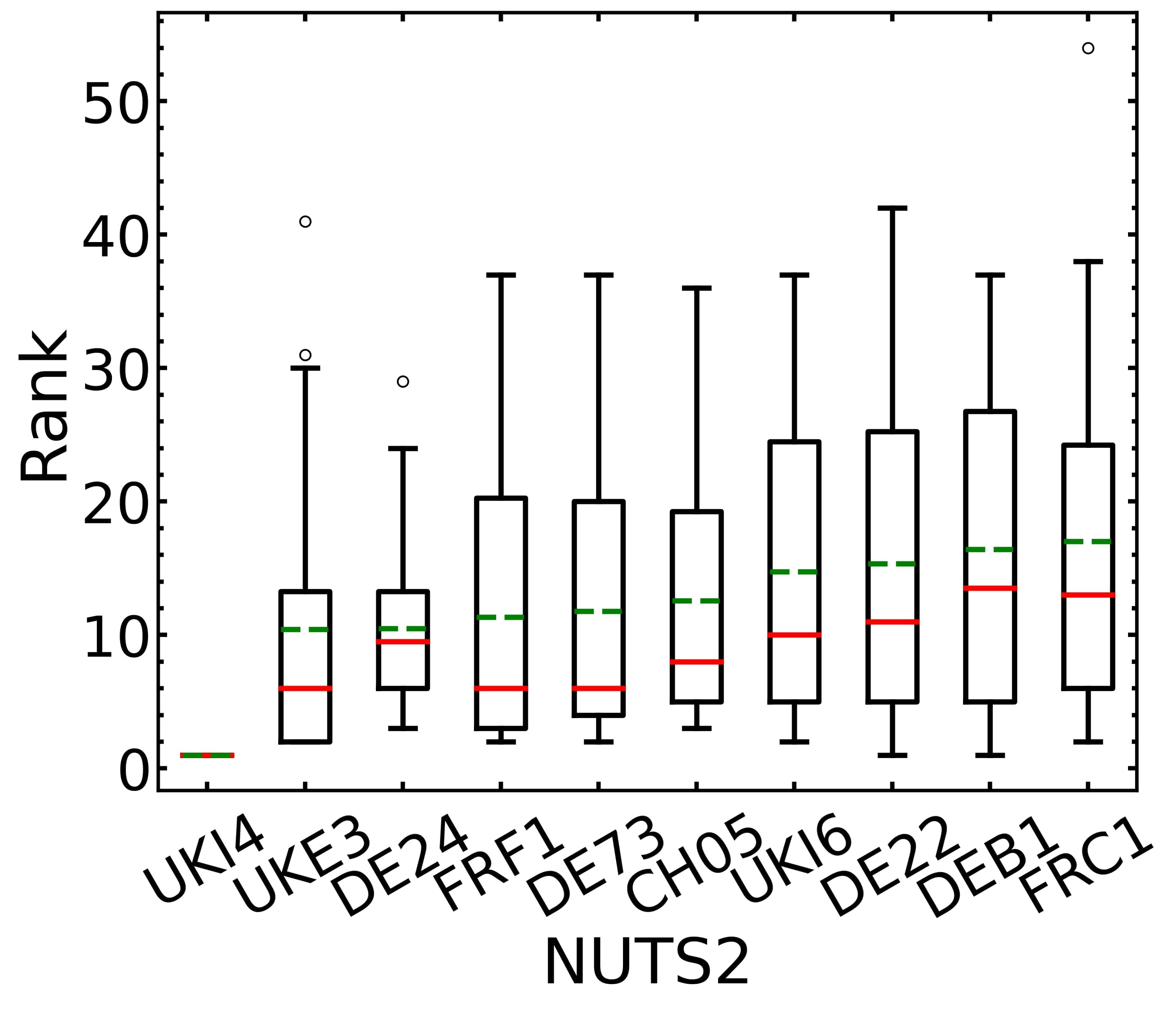} }}
		\subfloat[\label{fig:fig4b}]{{\includegraphics[width=0.45\textwidth]{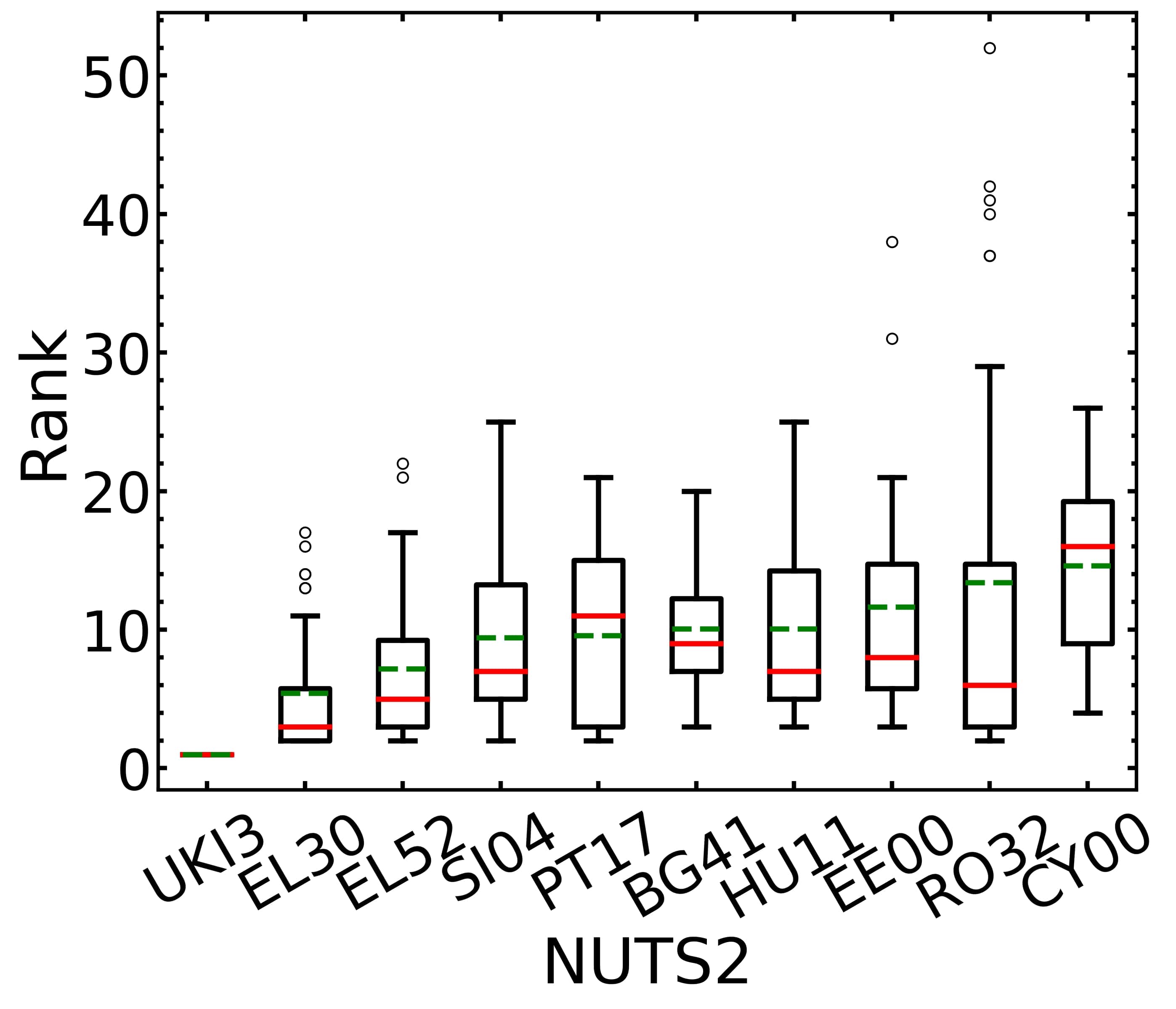} }}\\
		\subfloat[\label{fig:fig4c}]{{\includegraphics[width=0.45\textwidth]{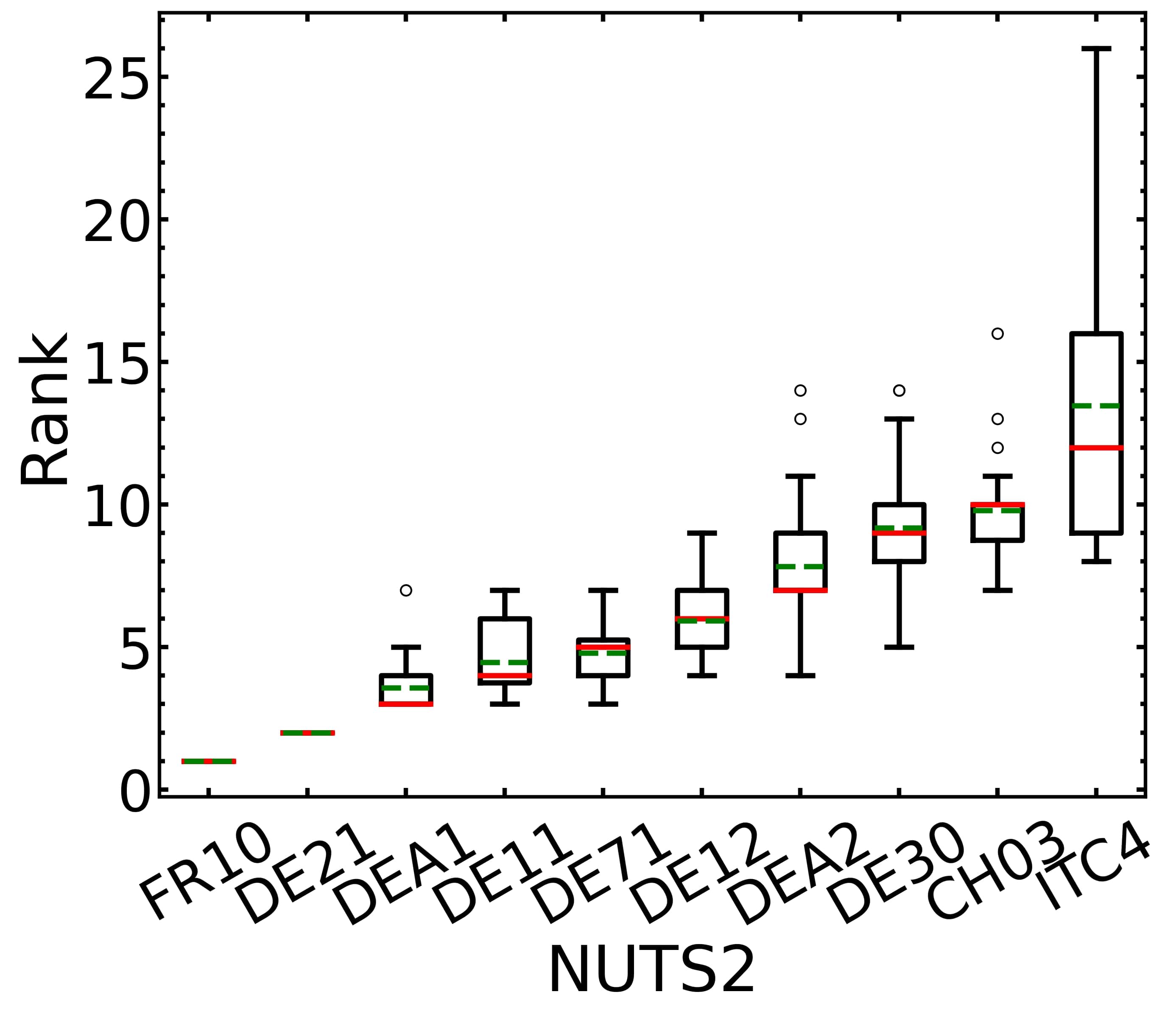} }}\\
		\caption{Boxplot of the top $10$ NUTS$2$ regions for a) patents, b) FPs, and c) common in terms of average ranking, according to the number of triangles ranking of the $28$ windows (green dashed lines). Red lines show the median value of their ranking for the $28$ windows.}
		\label{fig:fig4}
	\end{center}
\end{figure}

Finally, we present similar figures to the ones of the manuscript that have been created using the triangles number, instead of the clustering coefficient value. Fig. \ref{fig:fig4} shows the rank of top $10$ NUTS$2$ regions for patents, FPs and the common network. This ranking is made according to the average of the regions rank in all $28$ windows, based on the number of triangles that they are part of.

\begin{figure}[H]
	\begin{center}
		\subfloat[\label{fig:fig5a}]{{\includegraphics[width=0.45\textwidth]{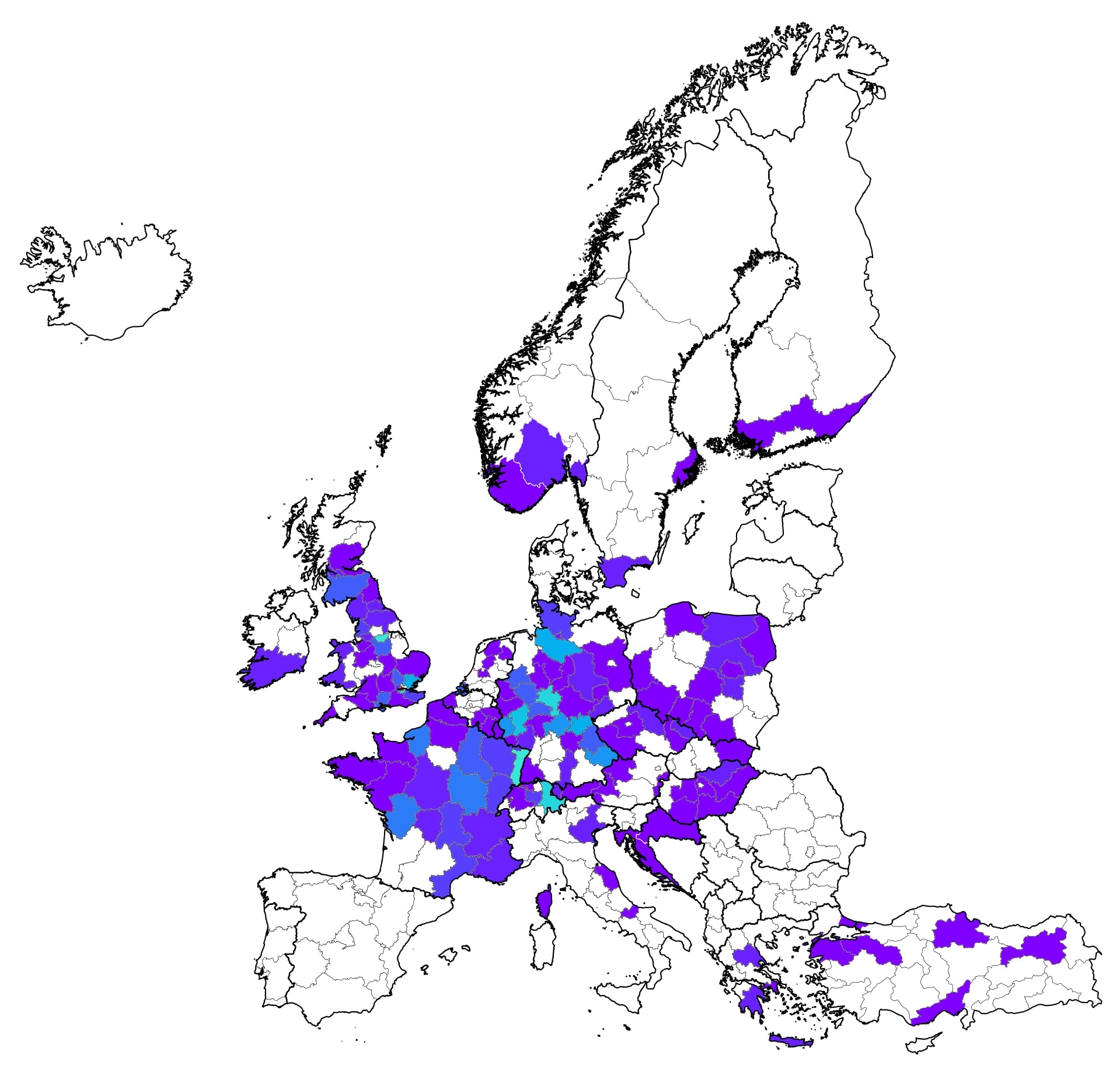} }}
		\subfloat[\label{fig:fig5b}]{{\includegraphics[width=0.45\textwidth]{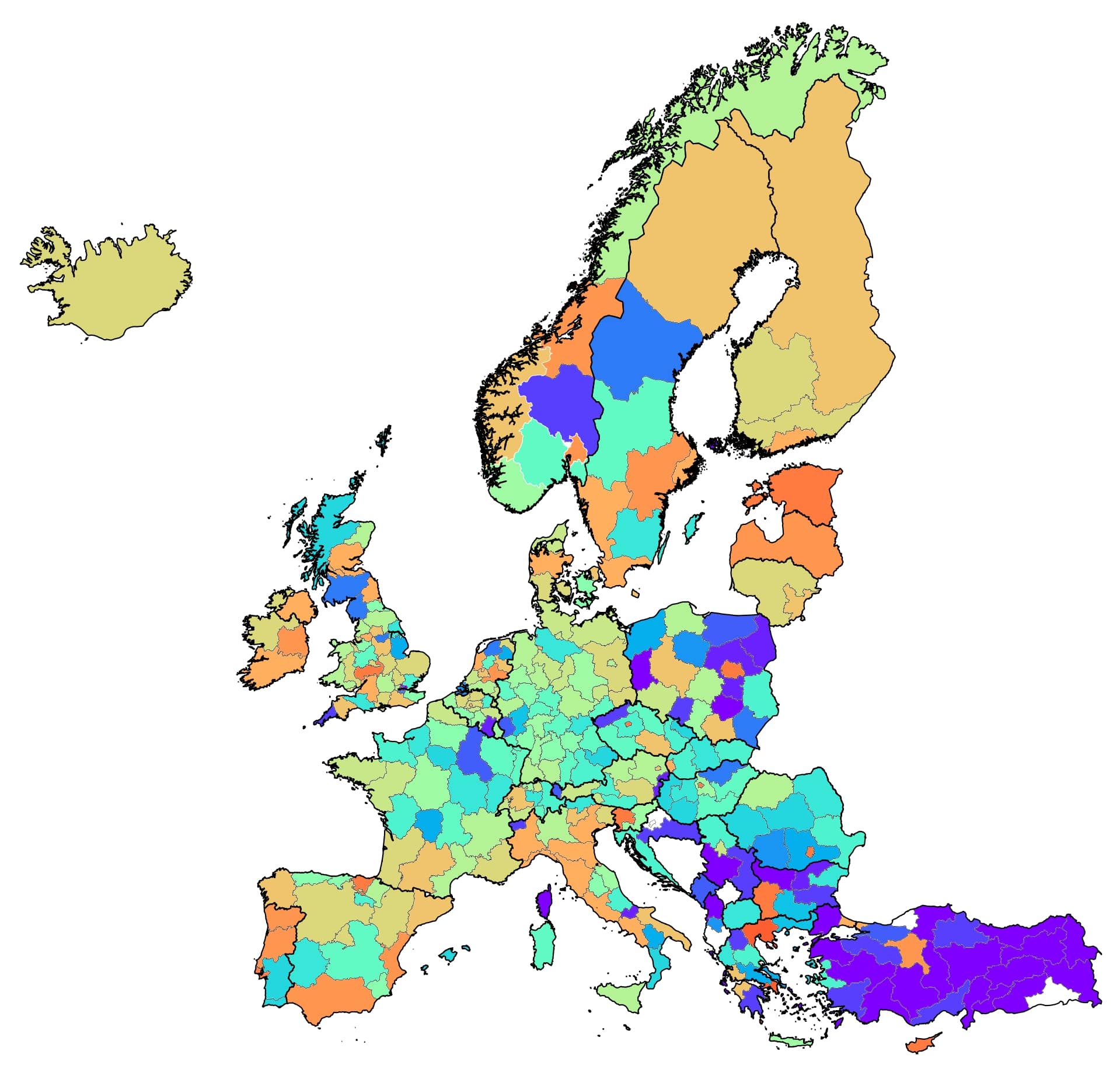} }} \\
		\subfloat[\label{fig:fig5c}]{{\includegraphics[width=0.45\textwidth]{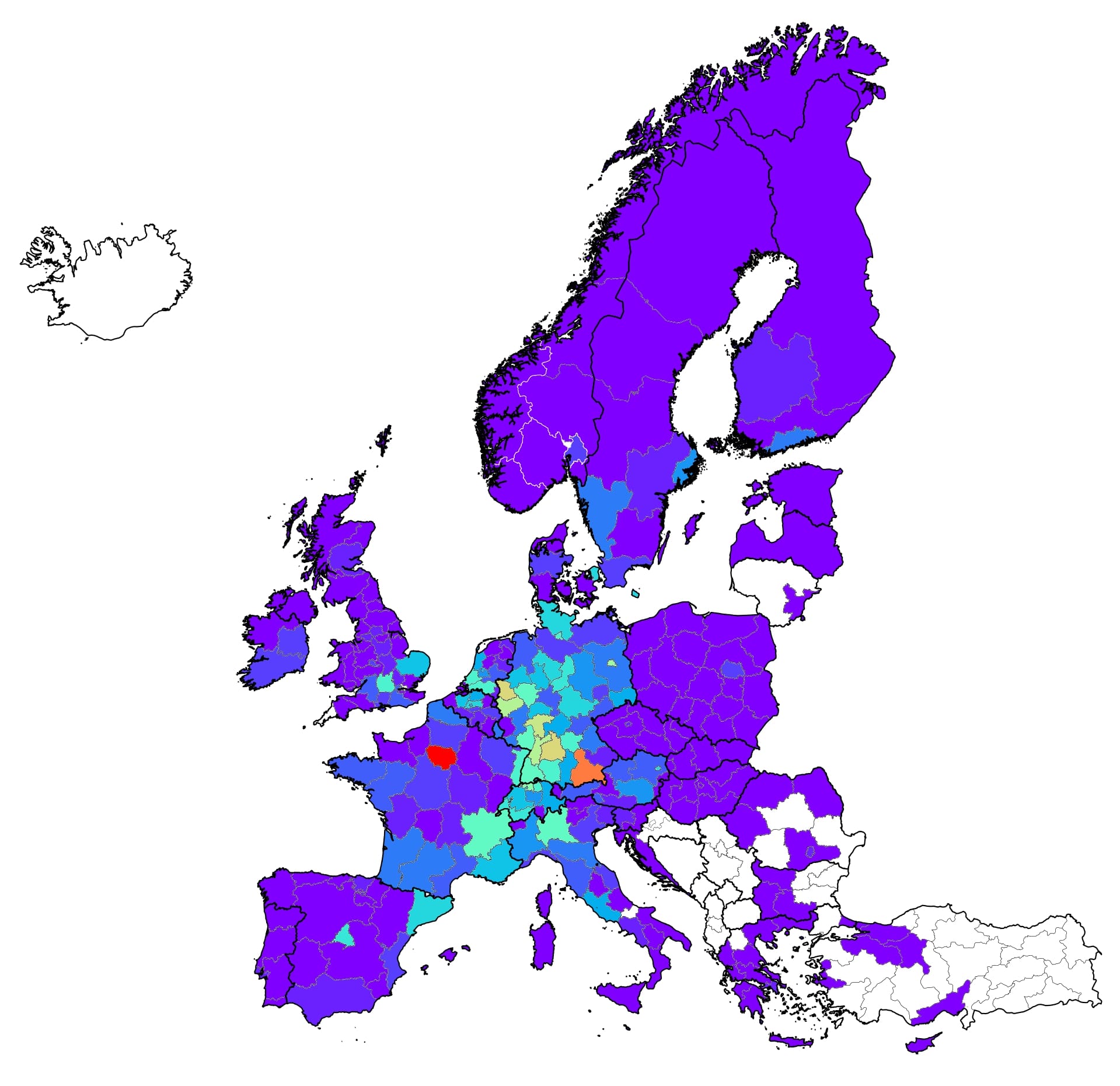} }}\\
		\subfloat[\label{fig:fig5d}]{{\includegraphics[width=1\textwidth]{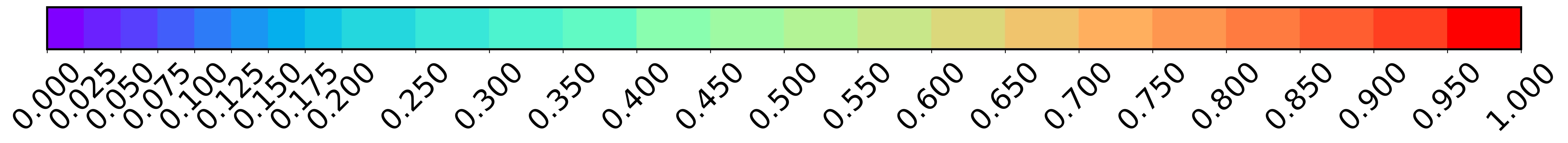} }}
		\caption{(Color online) European color-based map of the averaged number of triangles (normalized over the maximum number of triangles) of each NUTS$2$ region. Real data for a) patents, b) FPs, and c) common layer. As the average value of the normalized number of triangles increases, the colors change according to the d) colorbar (from left to right). White color indicates regions with zero triangles in that layer after the removal of the common triangles, or zero common triangles.}
		\label{fig:fig5}
	\end{center}
\end{figure}

Fig. \ref{fig:fig5}, shows the color-based European map for these three cases. These colors are according to the normalized average number of triangles of the $28$ windows. The regions with white color indicate cases where there are no triangles formed, or none remain after the removal of the common triangles (in the first two maps), or that there are no common triangles between the two layers (in the third one).

